\documentclass{scrartcl}
\usepackage{amsmath}
\areaset{16cm}{25cm}
\usepackage{graphicx} 
\usepackage{lineno}
\usepackage[margin=10pt,font=small,labelfont=bf,
labelsep=period,format=plain]{caption}
\usepackage{siunitx}
\usepackage{hyperref}
\usepackage{cite}
\usepackage[onehalfspacing]{setspace}
\usepackage{multicol}
\usepackage{hyphenat}
\usepackage{url}
\usepackage{upgreek}
\usepackage{xcolor}
\usepackage[resetlabels]{multibib}
\newcites{SI}{SI Appendix References}
\makeatletter
\renewcommand{\fnum@figure}{Fig. \thefigure}
\makeatother
\title{Depletion-Induced Interactions Modulate Nanoscale Protein Diffusion in Polymeric Crowder Solutions}

\author{
    \normalsize Michelle Dargasz $^{1\ast}$,
    Nimmi Das Anthuparambil$^{1}$,
    Sebastian Retzbach$^{2}$,
    Anita Girelli$^{3}$, \and
    \normalsize  Sonja Timmermann$^{1}$,
    Johannes Möller$^{4}$,
    Wonhyuk Jo$^{4}$,
    Aliaksandr Lenonau$^{1,4}$, \and
    \normalsize Agha Mohammad Raza$^{1}$,
    Maddalena Bin$^{3}$,
    Jaqueline Savelkouls$^{5}$,
    Iason Andronis$^{3}$,\and
    \normalsize Frederik Unger$^{1}$, 
    Felix Brausse$^{4}$,
    Jörg Hallmann$^{4}$,
    Ulrike Boesenberg$^{4}$,\and
    \normalsize Jan-Etienne Pudell$^{4}$,
    Angel Rodriguez-Fernandez$^{4}$,
    James Wrigley$^{4}$, \and
    \normalsize Roman Shayduk$^{4}$,
    Mohamed Youssef$^{4}$,
    Alexey Zozulya$^{4}$, 
    Anders Madsen$^{4}$, \and
    \normalsize Felix Lehmkühler$^{6,7}$,
    Fivos Perakis$^{3}$,
    Fajun Zhang$^{2}$,
    Frank Schreiber$^{2}$, \and 
    \normalsize Michael Paulus$^{5}$,
    Christian Gutt $^{1\dagger}$
}
  
\date{%
  \small
  $^{1}$Department Physik, Universität Siegen, Walter-Flex-Strasse 3, 57072 Siegen, Germany\\[0.5ex]
  $^{2}$Institut für Angewandte Physik, Universität Tübingen, Auf der Morgenstelle 10, 72076 Tübingen, Germany\\[0.5ex]
  $^{3}$Department of Physics, AlbaNova University Center, Stockholm University, 10691 Stockholm, Sweden\\[0.5ex]
  $^{4}$European X-Ray Free-Electron Laser Facility, Holzkoppel 4, 22869 Schenefeld, Germany\\[0.5ex]
  $^{5}$Fakultät Physik/DELTA, TU Dortmund, 44221 Dortmund, Germany\\[0.5ex]
  $^{6}$The Hamburg Centre for Ultrafast Imaging, Luruper Chaussee 149, 22761 Hamburg, Germany\\[0.5ex]
  $^{7}$Deutsches Elektronen-Synchrotron DESY, Notkestr. 85, 22607 Hamburg, Germany\\[2ex]
  $^\ast$Corresponding author. Email: michelle.dargasz@uni-siegen.de\\
  $^\dagger$Corresponding author. Email: christian.gutt@uni-siegen.de\\[2ex]
  Classification: Physical Science, Biophysics and Computational Biology\\[2ex]
}

\begin{document}

\maketitle
\newpage
\section*{Abstract}
Macromolecular crowding plays a crucial role in modulating protein dynamics in cellular and in vitro environments. Polymeric crowders such as dextran and Ficoll are known to induce entropic forces, including depletion interactions, that promote structural organization, but the nanoscale consequences for protein dynamics remain less well understood. Here, we employ megahertz X-ray photon correlation spectroscopy (MHz-XPCS) at the European XFEL to probe the dynamics of the protein ferritin in solutions containing sucrose, dextran, and Ficoll. We find that depletion-driven short-range attractions combined with long-range repulsions give rise to intermediate-range order (IRO) once the polysaccharide overlap concentration $c^*$ is exceeded. These IRO features fluctuate on microsecond to millisecond timescales, strongly modulating the collective dynamics of ferritin. The magnitude of these effects depends sensitively on crowder type, concentration, and molecular weight. Normalizing the crowder concentration by $c^*$ reveals scaling behavior in ferritin self-diffusion with a crossover near 2$c^*$, marking a transition from depletion-enhanced mobility to viscosity-dominated slowing. Our results demonstrate that bulk properties alone cannot account for protein dynamics in crowded solutions, underscoring the need to include polymer-specific interactions and depletion theory in models of crowded environments.

\section*{Significance statement}
Understanding how proteins move and interact in crowded environments is essential for unraveling the physical basis of cellular processes. Polymer-based crowders are widely used to mimic intracellular crowding. While the structural consequences of their modulation of interparticle interactions are well studied, their impact on protein dynamics remains poorly understood. We reveal that protein dynamics is strongly influenced by depletion interactions and transient nanoscale structures, with polymer concentration and molecular weight as key factors. Our findings highlight the need for refined crowding models incorporating polymer physics to better connect in vitro studies with cellular complexity.



\section*{Introduction}
Macromolecular crowding is a defining feature of the cellular environment \cite{ellis2001macromolecular,laurent1963interaction,minton1981excluded,minton1997influence}, where high concentrations of macromolecules significantly influence protein dynamics \cite{grimaldo2019dynamics,yu2016biomolecular,roosen2011protein,luby1999cytoarchitecture} - a key factor in processes such as molecular transport, signal transduction, and enzymatic activity \cite{heinen2012viscosity,dix2008crowding,gnutt2016macromolecular}. Numerous studies have shown that crowding can, for instance, modulate biochemical reactions by hindering protein diffusion or enhancing complex formation through entropic and enthalpic interactions \cite{homchaudhuri2006effect,derham2006effect,ren2003effects,wenner1999crowding}.

To mimic crowded conditions in vitro, studies commonly employ widely adopted standard crowders, such as polyethylene glycol (PEG), Ficoll, and dextran \cite{ranganathan2022ficoll,biswas2018mixed,rusinga2017soft}. In concentrated solutions, these flexible polymers undergo structural transitions, where polymer overlap and the formation of network structures characterized by their associated correlation lengths emerge as key descriptors on the nanoscale \cite{colby2010structure,teraoka2002polymer,hirata2003small}. Within such polymeric matrices, proteins often exhibit anomalous diffusion \cite{höfling2013anomalous,metzler2014anomalous,srinivasan2024breaking,szymanski2009elucidating,sabri2020elucidating}.

In protein-polymer mixtures non-adsorbing polymers induce depletion interactions, which are well-established as drivers of short-range attractive forces between proteins \cite{asakura1954interaction, vrij1976polymers, tuinier2011colloids,roth2010fundamental, roth2000depletion,bechinger1999understanding}. These interactions can facilitate protein-protein association, reversible aggregation, and liquid-liquid phase separation (LLPS) into protein-rich and protein-poor phases \cite{neu2006depletion,smith1995depletion,zhang2023depletion,gogelein2009polymer,tuinier2000depletion}. 
When short-range attractions are counterbalanced by long-range repulsions, proteins can form clusters or intermediate range order (IRO) on nanometer length scales \cite{von2019dynamic,liu2011lysozyme}. Such systems are of particular interest, as protein clustering is implicated in both physiological functions and pathological conditions, including neurodegenerative diseases such as Alzheimer’s and Parkinson’s \cite{piazza2006micro,kovalchuk2009formation}. 

While the resulting structural organization in protein systems with short-range attractions and long-range repulsions has been extensively characterized by scattering techniques, where features such as a low-$q$ peak in the static structure factor $S(q)$ are hallmarks of IRO \cite{liu2011lysozyme,godfrin2014generalized}, the dynamic consequences of these complex interactions remain poorly understood. Simulations by Riest \textit{et al.}~\cite{riest2015short} indicate that intermediate-range order (IRO) strongly affects dynamics and hydrodynamics compared with purely repulsive or purely attractive reference systems. However, experimental validation remains scarce, primarily due to the difficulty of resolving nanoscale protein dynamics on microsecond- to millisecond timescales -- a regime for which  megahertz X-ray photon correlation spectroscopy (MHz-XPCS) is ideally suited \cite{Reiser_natcomm_2022,Gir_arx_2024}.

To investigate nanoscale crowding -- specifically polymer overlap and depletion effects -- on collective protein dynamics and single-particle diffusion, we employed MHz-XPCS at the European XFEL to probe ferritin in solutions of sucrose, dextran, and Ficoll. Our experiments provide direct evidence that increasing crowder concentration substantially modifies collective diffusion through depletion-driven, transient IRO. This structural order fluctuates on microsecond-millisecond timescales and, for example, reduces hydrodynamic interactions, resulting in enhanced sedimentation coefficients. Our measurements enable quantification of depletion attraction strength and allow estimation of the lifetimes of protein complexes stabilized by IRO.

Importantly, we find that ferritin self-diffusion displays a non-monotonic dependence on crowder concentration, governed by polymer overlap concentration and depletion layer thickness. These deviations from simple scaling -- resembling those observed for nanoparticles in PEG solutions \cite{kohli2012diffusion} -- emphasize the role of nanoscale effects that can be directly resolved by XPCS.  

Taken together, our results demonstrate that  crowders profoundly shape protein dynamics through emergent structural organization and modifications of the local interaction landscape. This highlights the need for a more nuanced view on protein dynamics in crowded environments -- one that accounts for depletion interactions, polymer network structure, and transient nanoscale order as key contributors to protein mobility and phase behavior in complex environments.\\

\section*{Results}
\textbf{Modeling a crowded environment utilizing crowding agents.} 
In this study, we investigated the protein ferritin, dissolved in solutions containing various macromolecular crowders. Ferritin is a spherical protein complex composed of 24 subunits that assemble into a hollow shell capable of storing up to 4,500 Fe(III) atoms \cite{Theil_wiley_1990, Har_bba_1996}, thus fulfilling its primary function as iron storage \cite{Cha_structuralbio_1999}. Its hydrodynamic radius has been determined to be $R_{\text{h}} = 6.85$ nm \cite{Moh_biomacrom_2021}. Due to its monodispersity and stability over a broad range of pH and temperature conditions, ferritin is also widely used in vaccine development \cite{Rod_pharma_2021} and drug delivery applications \cite{Khos_contrrelease_2018, Luc_Interbiomacrom_2022}. These attributes, combined with the strong X-ray scattering contrast provided by its iron-rich core, make ferritin an ideal globular model protein for X-ray scattering studies under crowded conditions. Several standard crowding agents of varying molecular sizes were used, including sucrose (M$_{\mathrm{w}}$= 342.3 g/mol) and the polysaccharides Ficoll400 (M$_{\mathrm{w}}$= 400 kg/mol) and dextran with three different molecular weights (M$_{\mathrm{w}}$= 40, 100, 500 kg/mol; in the following referred to as dextran 40, dextran 100 and dextran 500). 

\textbf{Accessing nanoscale protein dynamics by extracting intensity autocorrelation functions utilizing XPCS.} 
We investigated the influence of standard crowders on protein dynamics via MHz-XPCS experiments at the Materials Imaging and Dynamics (MID) instrument of the European XFEL \cite{madsen2021materials}. This technique enables direct access to collective and self-diffusion on nanometer length scales and microsecond timescales, even in highly concentrated solutions \cite{anthuparambil2024salt,anthuparambil2023exploring,timmermann2023x,girelli2021microscopic,begam2021kinetics}. Moreover, XPCS allows determination of the hydrodynamic function, aiding in the study of long-range many-body interactions mediated by the solvent \cite{dallari2021microsecond}. These capabilities make MHz-XPCS a powerful tool for investigating molecular crowding effects. Its successful application to protein dynamics has been demonstrated by Reiser \textit{et al.} \cite{Reiser_natcomm_2022}, Girelli \textit{et al.} \cite{Gir_arx_2024} and Anthuparambil \textit{et al.} \cite{anthuparambil2025softness}.

\begin{figure}[h!]
\centering
\captionsetup{width=\linewidth}
\includegraphics[scale=.72]{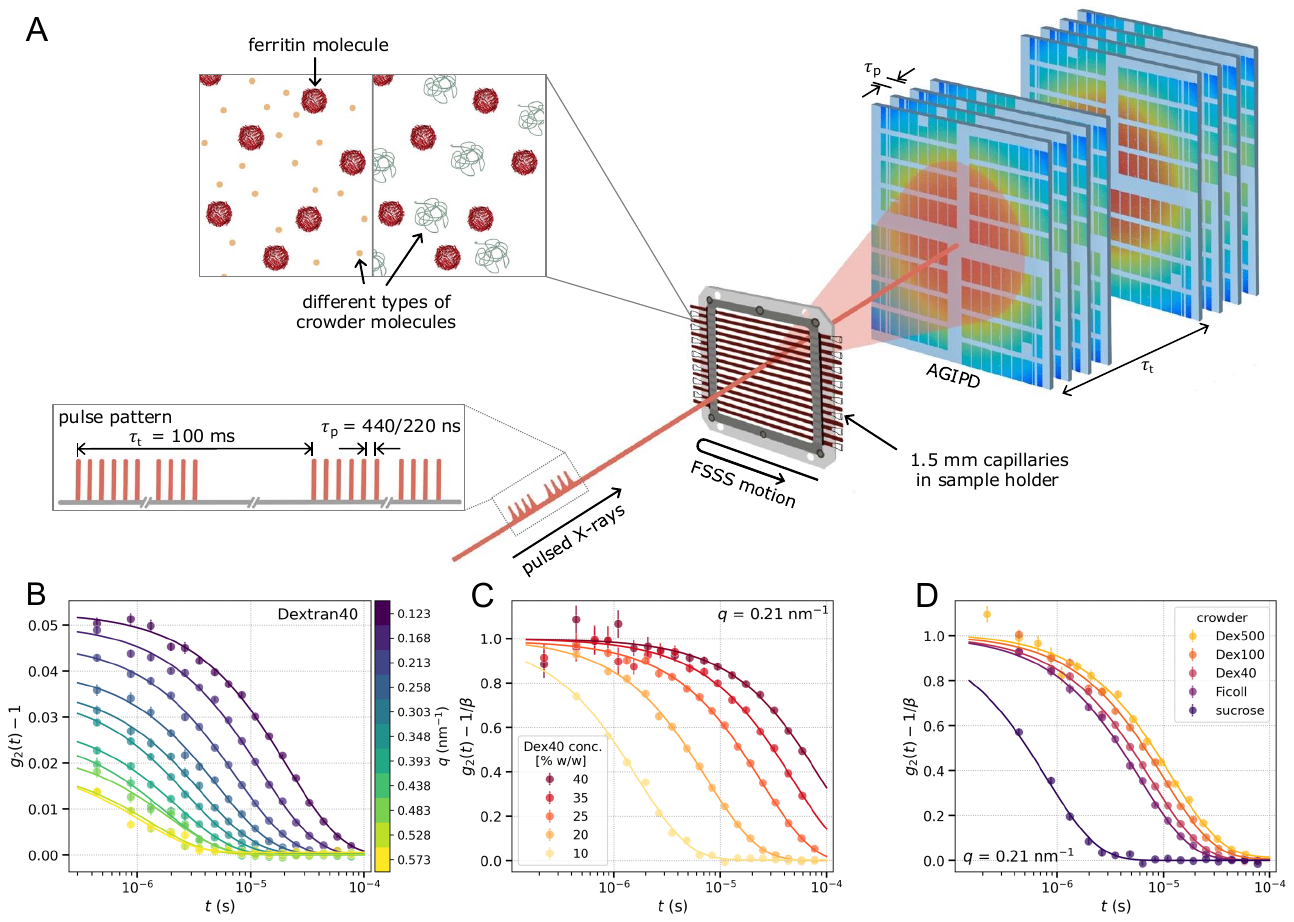}
\caption{MHz-XPCS measurements at the MID instrument of the European XFEL. (A) Schematic of the experimental setup. Coherent X-ray pulses illuminate ferritin solutions containing different crowder types and concentrations, producing speckle patterns recorded by the AGIPD. A new pulse train is delivered every 100 ms, with intra-pulse spacings of 440 ns and 220 ns used. The Fast Solid Sample Scanner (FSSS) moves the sample between trains, ensuring that each train probes a fresh spot. Intensity fluctuations within a train are analyzed to extract dynamic information by calculating intensity autocorrelation functions ($g_2(t)$). (B) $q$-dependent $g_2(t)$ functions for ferritin in a 25 \%w/w dextran 40 solution. (C) $g_2(t)$ functions at a fixed $q$-value of 0.21 nm$^{-1}$ for dextran 40 concentrations ranging from 10 to 40 \%w/w. (D) $g_2(t)$ functions at the same $q$-value for 20 \%w/w of dextran, Ficoll, and sucrose. Solid lines represent fits based on Eq. \ref{eq:1}. Error bar determination is described in Materials and Methods. The ferritin schematic is adapted from PDB ID:2w0o \cite{Val_JInorgBiochem_2012}.}
\label{fig:g2}
\end{figure}

Samples were loaded into quartz-glass capillaries and sealed with epoxy glue. The experimental setup for MHz-XPCS experiments at MID is shown in Fig.~\ref{fig:g2}A. Results are reported from two experimental campaigns using photon energies of 9 and 10 keV and beam sizes of 14.5 and 14.2 µm, respectively. X-ray pulse trains containing  200 respective 310 pulses with a spacing of 440 and 220 ns interacted with the sample, and pulse-resolved scattering patterns were recorded with the Adaptive Gain Integrating Pixel Detector (AGIPD), positioned at {7.15 m} and {7.68 m} from the sample, respectively. This configuration provided access to wavevectors $q$ in the range of 0.1 – 1 nm$^{-1}$ as well as 0.075 – 1 nm$^{-1}$ and enabled measurements on timescales from 0.44 to 88 µs and 0.22 to 69 µs. 

To reduce radiation damage and beam-induced heating, the X-ray intensity was strongly attenuated, resulting in photon fluxes of $10^{7} - 10^{8}$ photons per pulse incident on the sample. The sample volume was refreshed for each pulse train, ensuring that the total X-ray dose and dose rate remained below established damage thresholds \cite{Gir_arx_2024,anthuparambil2025softness}. Statistically robust data were obtained by averaging scattering patterns and autocorrelation functions from more than 12,000 distinct positions across multiple capillaries. In addition to the MHz-XPCS experiments, complementary small-angle X-ray scattering (SAXS) measurements were performed at the DELTA synchrotron radiation facility at TU Dortmund \cite{dargasz2022x}.


The observed dynamics can be attributed exclusively to ferritin, as no correlation and thus no dynamics were detected in the two-time correlation functions of the pure crowder solutions (\textit{SI Appendix}, Fig.~\ref{fig:ttc}). 
Protein dynamics were analyzed by calculating intensity autocorrelation functions, $g_2(q, t)$, from the temporal intensity fluctuations of the speckle patterns recorded within individual X-ray pulse trains. Fig.~\ref{fig:g2}B presents an example of the resulting $q$-dependent $g_2$($q$,$t$) functions, where $q$ refers to the momentum transfer as $q = 4\pi/\lambda$ sin($\theta$) with 2$\theta$ denoting the scattering angle and $\lambda$ indicating the X-ray wavelength. All $g_2$($q$,$t$) functions were collected at room temperature using a constant ferritin concentration of 55 mg/ml and varying crowder concentrations of 10 up to 40 \%w/w. 
The solid lines represent fits using a Kohlrausch-William-Watts (KWW) function \cite{Willxpcstheory}:
\begin{equation}
    g_2 (q,t) = 1 + \beta(q) \,  \mathrm{exp}[-2\,(\Gamma(q)\,t)^{\alpha}] \label{eq:1},
\end{equation}
where $\beta$($q$) denotes the $q$-dependent speckle contrast, modeled following \cite{Hru_PhysRevLett_2012,Leh_pnas_2020}, $\Gamma$($q$) is the $q$-dependent relaxation rate, and $\alpha$ is the KWW exponent. All correlation functions are well described with $\alpha=0.9$ indicating slightly sub-diffusive behavior, consistent with previous observations for the diffusion of nanoparticles, proteins, and fluorescent tracers in solutions containing soft polymers \cite{omari2009diffusion, banks2005anomalous, sanabria2007multiple}. Further details on extracting the $g_2$($q$,$t$) are provided in Materials and Methods. 

Fig.~\ref{fig:g2}C shows the $g_2$($t$) functions of ferritin in dextran 40 solutions at varying concentrations. 
Fig.~\ref{fig:g2}D compares ferritin dynamics in solutions containing different crowders at 20 \%w/w. In sucrose solutions, Ferritin exhibits significantly faster dynamics than in polysaccharide-based solutions, primarily due to differences in solution viscosity \cite{mazurkiewicz1998viscosity,masuelli2014dextrans}.
\begin{figure}[h]
\centering
\includegraphics[scale=.91]{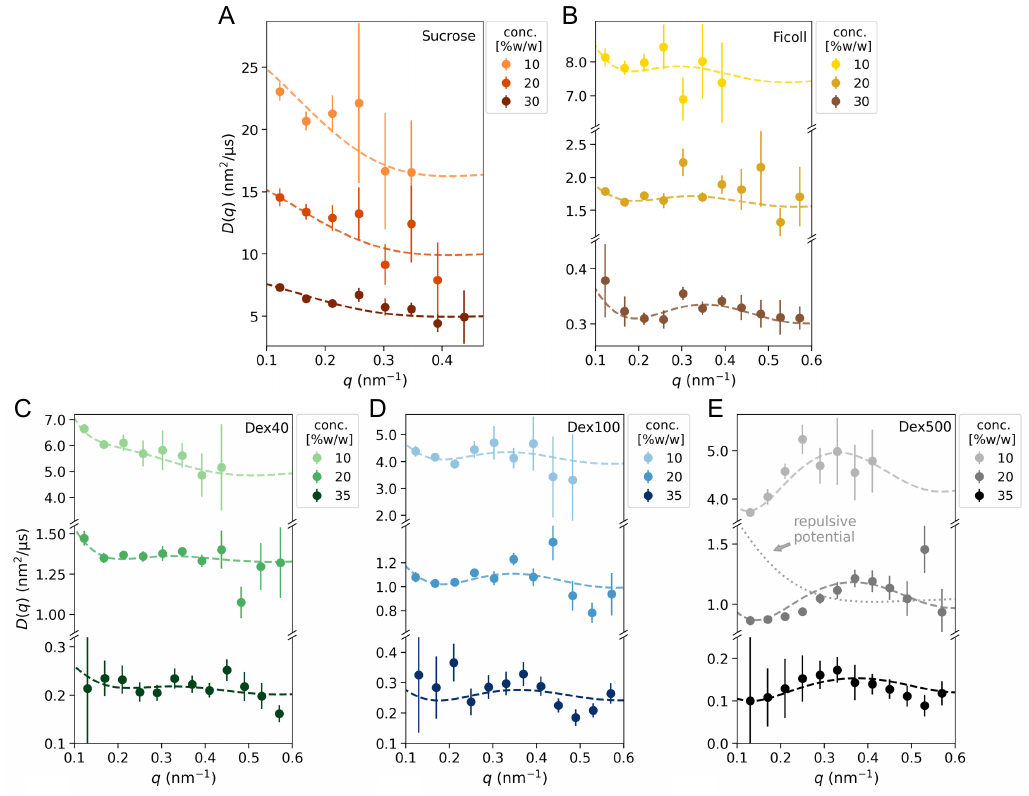}
\caption{Extracted $q$-dependent collective diffusion coefficients $D(q)= \Gamma(q)/q^2$ of ferritin in crowded solutions. (A)-(E) $D(q)$ for ferritin in solution containing sucrose (10, 20, 30 \%w/w), Ficoll400 (10, 20, 30 \%w/w), dextran 40 (10, 20, 35 \%w/w), dextran 100 (10, 20, 35 \%w/w), and dextran 500 (10, 20, 35 \%w/w). Error bars were obtained from least-squares fits of the $g_2$-functions. Dashed lines represent model calculations of $D$($q$) using (A)  a single-Yukawa potential and (B-E) a two-Yukawa potential (see Materials and Methods). In panel (E), the dotted line contrasts the observed $D(q)$ for polysaccharides with the $D(q)$ of a purely repulsive system, highlighting the presence of attractive forces in polysaccharide solutions.}
\label{fig:d(q)}
\end{figure}

\textbf{Collective protein dynamics mirrors interparticle interaction potential characteristics revealing a depletion-induced IRO in polysaccharide solutions. }
In crowded solutions at high particle concentrations, protein dynamics are governed  by both direct protein-protein interactions and indirect hydrodynamic interactions mediated by the solvent \cite{Lehmxpcstheory}. These interactions give rise to a wavevector-dependent collective diffusion coefficient, $D(q)$, which is extracted from exponential fits of the intensity autocorrelation functions, \( g_2 \), via  $D(q) = \Gamma(q)/q^2$. 

Fig.~\ref{fig:d(q)}A displays the resulting $D(q)$ for ferritin in solution with sucrose, a small non-polymeric crowder. With increasing $q$, $D(q)$ decreases and reaches a minimum at wavevectors corresponding to the maximum of the static structure factor $S(q)$, reflecting the inverse relation between $D(q)$ and $S(q)$. The characteristic shape of $D(q)$ indicates predominantly repulsive protein interactions, in agreement with recent measurements of collective ferritin diffusion in aqueous solutions \cite{Gir_arx_2024}.
The collective dynamics of ferritin in polysaccharide solutions are presented in Fig.~\ref{fig:d(q)}B-E. As for ferritin in sucrose, the minimum around $q=0.55$ nm$^{-1}$ coincides with the maximum of the structure factor. However, in contrast to sucrose, increasing the concentration of polymeric crowders results in a pronounced modulation in the slope of the $D(q)$ at low $q$, leading to a deviation from the monotonic decrease in $D(q)$ characteristic of repulsive systems (dotted line in Fig.~\ref{fig:d(q)}E). This behavior points to the coexistence of short-range attractions and long-range repulsions, consistent with the development of IRO.

IRO is characterized by spatial correlations extending over intermediate length scales -- typically two to three protein diameters -- arising from the competition of  interactions rather than from stable aggregation or clustering \cite{zhang2024effective, liu2011lysozyme}. These correlations correspond to transient complexes in which proteins dynamically exchange between monomeric and associated states \cite{liu2011lysozyme}. As a result, the IRO remains dynamic, with structural fluctuations occurring on the microsecond timescale, as revealed by our XPCS measurements (see below).

This intermediate-range structural organization reduces protein mobility at length scales corresponding to $q$-values two to three times smaller than those associated with the structure factor maximum, i.e. the typical ferritin monomer-monomer peak position. Similar IRO formation has been reported for concentrated lysozyme \cite{Por_PhysChemLett_2010,Liu_PhysChemB_2011} and monoclonal antibody (mAb) solutions \cite{Yea_biophys_2014}, where enhanced attractive interactions arise either from high protein concentrations or from specific interprotein interactions. 

In systems containing polymer-like crowders, depletion interactions provide a well-established mechanism to induce short-range attraction. As polymers lose configurational entropy near protein surfaces, polymer-depleted regions form around the particles \cite{asakura1954interaction, vrij1976polymers, tuinier2011colloids}. When these regions overlap, the excluded volume for the polymers is reduced, thereby increasing the system’s entropy and generating an effective attractive force between proteins. 

In addition to Coulomb repulsion (partially screened by salt ions), polymers themselves can give rise to long-range repulsive interactions. This so-called free polymer-induced repulsion emerges from the accumulation of flexible polymer chain ends near protein surfaces, which produces an entropic repulsion between particles \cite{semenov2008theory,shvets2013effective,semenov2015theory}.

The transition of the $D(q)$ profile from a monotonic decrease to a distinct slope modulation at low $q$ occurs when the polymer overlap concentration $c^*$ is exceeded (compare with Fig.~\ref{fig:d(q)}C, where $c^*_{\text{dex40}}$ = 12.75 \%w/w). The overlap concentration marks the onset of the semi-dilute regime and was determined by concentration-dependent macroscopic viscosity measurements (\textit{SI Appendix}, Table \ref{table:overlap}). IRO emerges above $c^*$ not only because depletion attraction increases with crowder concentration \cite{gogelein2008phase}, but also because long-range free-polymer induced repulsion can develop only in the semi-dilute regime \cite{shvets2013effective}.

\textbf{Extracting structural and hydrodynamic information from protein dynamics.} The dashed lines in Fig.~\ref{fig:d(q)} represent model calculations of $D(q)$, providing insight into the static structure and hydrodynamic interactions within the solution. The approach relies on the relation between $D(q)$, the static structure factor $S(q)$, the hydrodynamic function $H(q)$, and $D_0(c_{\text{cr}})$, the diffusion coefficient of the protein in crowder solutions in absence of  interactions \cite{nagele1996dynamics}
\begin{equation}
    D(q) = D_0(c_{\text{cr}}) \cdot \frac{H(q)}{S(q)} \label{eq:D(q)}.
\end{equation}
Within this framework, $H(q)/S(q)$ was computed from the underlying particle-particle interaction potentials, as described in the \textit{SI Appendix}. To capture both the short-range attraction and long-range repulsion present in polysaccharide-containing solutions, a two-Yukawa potential was employed, a model commonly used to describe systems exhibiting IRO \cite{Liu_PhysChemB_2011}. The reduced form of this potential is given in \textit{SI Appendix}, Eq.~\ref{eq:two-yukawa}.
For sucrose solutions, where repulsive interactions dominate, a single-Yukawa potential was sufficient. The hydrodynamic function was obtained using the $\delta\gamma$ expansion of Benakker and Mazur \cite{Bee_PhysAStat_1984, Bee_PhysAStat_1983} (see \textit{SI Appendix} for details). Optimizing the parameters of the interaction potentials enabled simultaneous determination of $S(q)$ and $H(q)$. The corresponding potentials are shown in \textit{SI Appendix}, Fig.~\ref{fig:pot}, and the fitted parameters are listed in Table~\ref{table:params}. Despite the empirical nature of this modeling approach models and the simplicity of the assumed potentials, the calculated $D(q)$ agrees reasonably well with the experimental data (compare with Fig.~\ref{fig:d(q)}). 



\begin{figure}[h!]
\centering
\includegraphics[scale=.48]{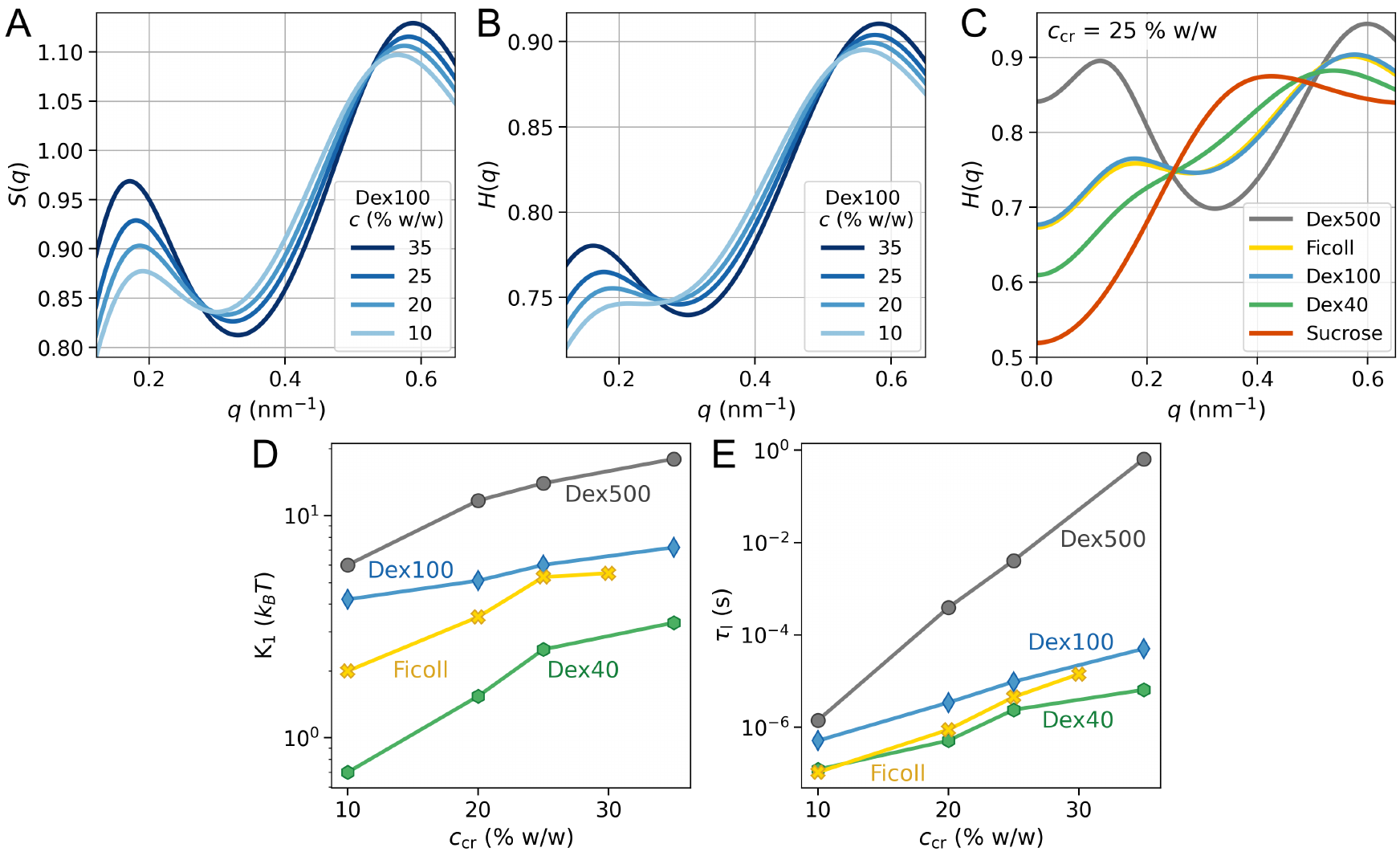}
\caption{Effects of concentration and molecular weight on structure formation due to IRO. (A) Modeled static structure factor $S(q)$ and (B) hydrodynamic function $H(q)$ based on a two-Yukawa potential for ferritin in dextran 100 solution at 10, 20, 25 and 35 \%w/w. (C) $H(q)$ for ferritin in solutions with different crowders at $c_{\text{cr}}= 25 $ \%w/w. (D) Attractive potential strength parameter $K_1$, obtained by fitting the experimental $D(q)$ data with Eq.~\ref{eq:D(q)} using a two-Yukawa potential. (E) Estimated lifetime of ferritin complexes stabilized by IRO in polysaccharide solutions, calculated from Eq.~\ref{eq:lifetime}, as a function of crowder concentration.}
\label{fig:s(q)}
\end{figure}
\noindent \textbf{Depletion attraction scales with crowder concentration and molecular weight.} 
The concentration-dependent structure factor $S(q)$ (Fig.~\ref{fig:s(q)}A) and hydrodynamic function $H(q)$ (Fig.~\ref{fig:s(q)}\linebreak B) for ferritin in dextran 100 solutions both demonstrate that IRO becomes increasingly pronounced with rising concentration, as reflected by the progressive enhancement of the low-$q$ contribution. Since $H(q) = 1$ corresponds to the absence of hydrodynamic interactions \cite{dallari2021microsecond}, deviations toward lower values quantify the strength of hydrodynamic hindrance. Thus, the observed increase in $H(q)$ at $q= 0.19$ $\text{nm}^{-1}$ indicates a reduction in hydrodynamic hindrance associated with IRO formation. The close relationship between $S(q)$ and $H(q)$ is well known and reflects the reduced hydrodynamic drag in systems where correlated particle motion leads to more coherent flow patterns and less mutual interference.

The hydrodynamic function also provides access to the sedimentation coefficient, since \linebreak $H(q \rightarrow 0)=K=V_{\text{sed}}/V_0$, where $V_{\text{sed}}$ is the mean sedimentation velocity in a concentrated dispersion and $V_0$ the single-particle sedimentation velocity \cite{riest2015short}. 
Sedimentation is an essential mode of mass transport, playing a central role in the intracellular distribution of biomolecules \cite{li2011analysis, hinderliter2010isdd} and in regulating phase separation.
Figure~\ref{fig:s(q)}C shows $H(q)$ for ferritin in 25 \%w/w solutions of various crowders. Compared with sucrose solutions, where repulsive interactions dominate, polysaccharide solutions such as dextran 500 show enhanced sedimentation coefficients $K$. This increase indicates that short-range attraction introduced by macromolecular crowding can promote more efficient transport processes and potentially facilitate phase separation.

The observed dependence of structure and dynamics on both molecular weight and crowder concentration correlates with an enhanced depletion attraction, as evidenced by the increasing attractive potential strength parameter $K_1$ (Fig.~\ref{fig:s(q)}D), and the deepening interaction potential minimum (\textit{SI Appendix}, Fig.~\ref{fig:pot}). Theoretical calculations of depletion-induced interaction potentials based on refs. \cite{ilett1995phase,chatterjee1998microscopic} are consistent with these experimental trends. Following the approach of \cite{Car_PhysChemB_2011}, the potentials further allow an approximate estimation of the lifetimes of protein complexes stabilized by IRO, by considering the characteristic escape time of two particles from a potential well. For the rather shallow potentials to be considered here, we make use of an expression for the lifetime derived in \cite{abkenar2017dissociation}: 
\begin{align}
\tau_{\text{l}} = \frac{\Delta^2}{D_{\text{S}}^{\text{H}}(c_{\text{cr}})} \,\frac{1}{U^2}\left(\mathrm{exp}\left(U\right)-1\right)-\frac{1}{U} \label{eq:lifetime}
\end{align}
where $\Delta \sim 0.1 \sigma$, with $\sigma = 11.8$ nm corresponding to the ferritin diameter \cite{Wie_JApplCryst_2021} and $U$ denotes the minimum of the potential in units $k_{\text{B}}T$. $D_{\text{S}}^{\text{H}}(c_{\text{cr}})$ is the self-diffusion coefficient of ferritin in the presence of hydrodynamic interactions mediated by polysaccharides and water, for varying crowder concentrations. This parameter is derived from the hydrodynamic function $H(q)$ in the high-$q$ region, where $H(q\rightarrow \infty) = \frac{D_{\text{S}}^{\text{H}}(c_{\text{cr}})}{D_0(c_{\text{cr}})}$. Both $H(q\rightarrow \infty)$ and $D_0(c_{\text{cr}})$ were obtained by modeling the measured $D(q)$. The resulting estimates based on our fitted potential parameters are shown in Fig.~\ref{fig:s(q)}E. Higher dextran molecular weights lead to stronger depletion attraction and thus longer complex lifetimes. We note that the lifetimes estimated based on the interaction potential are consistent with the relaxation times measured via MHz-XPCS. In particular for dextran 500 at low $q$,  the correlation functions do not fully decay within the available time window (see \textit{SI Appendix}, Fig.~\ref{fig:g2_dex500}), indicating rather slow fluctuations of the IRO.

This observation carries important implications for biological systems. Although macromolecular crowding is generally associated with reduced association rate constants due to slowed diffusion, this effect can be offset by enhanced  intermolecular attraction -- such as depletion-induced forces -- which can accelerate binding \cite{berezhkovskii2016theory, Zos_pnas_2020, zhou2008macromolecular}. The pronounced increase in attractive interactions observed here under crowded conditions could therefore strongly influence protein-protein association dynamics, including processes such as enzymatic catalysis and cytokine-mediated immune signaling  \cite{schreiber2009fundamental}. Because both the strength of attraction and the lifetimes of associated complexes scale with crowder molecular weight and concentration, our findings may help reconcile the diverse and sometimes contradictory effects of crowding reported in the literature \cite{kuznetsova2014macromolecular, mittal2015macromolecular}.

\begin{figure}[h]
\centering
\includegraphics[scale=.47]{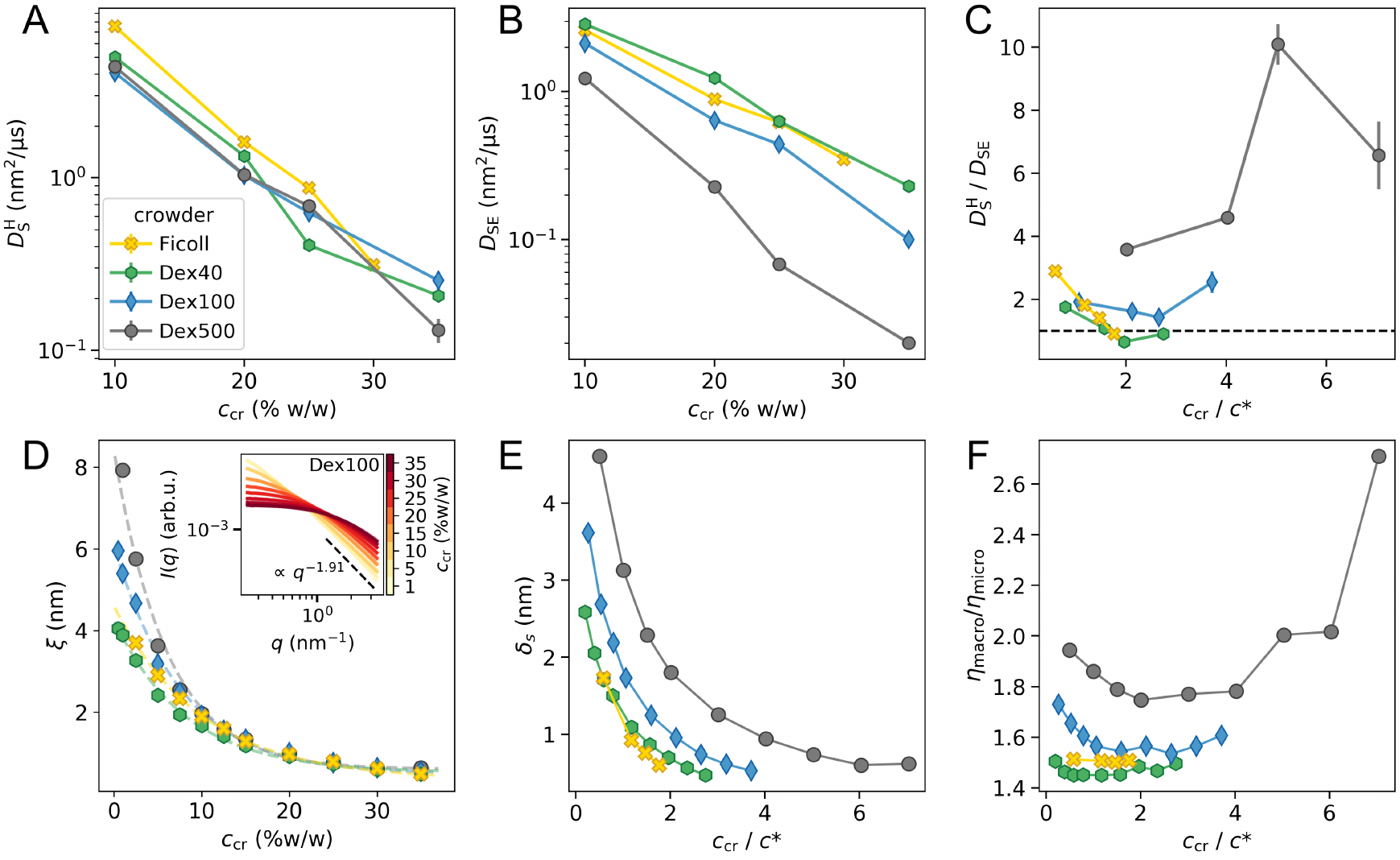}
\caption{Polysaccharide concentration and depletion effects on ferritin self-diffusion. (A) Microscopic self-diffusion coefficient $D_{\text{S}}^{\text{H}}(c_{\text{cr}})$ and (B) macroscopic Stokes-Einstein self-diffusion coefficient $D_{\text{SE}}(c_{\text{cr}})$ as a function of polysaccharide concentration $c_{\text{cr}}$. Error bars, obtained from error propagation of $D(q)$ fits and viscosity uncertainties, are smaller than symbols when not visible. (C) Ratio of microscopic to macroscopic self-diffusion coefficients as a function of crowder concentration normalized by overlap concentration value $c^*$. Error bars result from error propagation. (D) Correlation length $\xi(c_{\text{cr}})$ of crowder solutions obtained by fitting SAXS intensities of pure crowder solutions with Eq.~\ref{eq:O-Z}, dashed lines serve as visual guide. The inset shows example SAXS data for dextran 100. (E) Depletion layer thickness $\delta_{\text{s}}$ calculated using  Eq.~\ref{eq:depl_thick} (\textit{SI Appendix}). (F) Ratio of macroscopic to microscopic viscosity as a function of normalized crowder concentration ($c_{\text{cr}}/c^*$). Microscopic viscosity values were calculated using  Eq.~\ref{eq:micro_visc} (\textit{SI Appendix}).}
\label{fig:Dc}
\end{figure}

\textbf{Self-diffusion is strongly influenced by micro- and macroscopic viscosity variations arising from depletion effects.} To assess the impact of crowders on single-protein diffusion, the microscopic self-diffusion coefficient of ferritin obtained from XPCS, $D_{\text{S}}^{\text{H}}(c_{\text{cr}})$, was compared with the macroscopic diffusion coefficient $D_{\text{SE}}(c_{\text{cr}})$. The latter was calculated using the Stokes-Einstein equation, based on the experimentally determined bulk viscosities of the crowder solutions and the hydrodynamic radius of ferritin. The concentration dependence of $D_{\text{SE}}(c_{\text{cr}})$ is shown in Fig.~\ref{fig:Dc}B, revealing a clear molecular weight dependence. In contrast, $D_{\text{S}}^{\text{H}}(c_{\text{cr}})$ exhibits no pronounced molecular weight dependence (Fig.~\ref{fig:Dc}A), suggesting that in dense solutions, local polymer structure rather than overall chain length dominates in regulating protein mobility. 

To further probe the discrepancy between $D_{\text{S}}^{\text{H}}(c_{\text{cr}})$ and $D_{\text{SE}}(c_{\text{cr}})$, their ratio was analyzed as a function of normalized crowder concentration $c/c^*$ (Fig.~\ref{fig:Dc}C). The results show that $D_{\text{S}}^{\text{H}}/D_{\text{SE}}$ does not scale with bulk viscosity and displays a non-monotonic concentration dependence. At high crowder concentrations, $D_{\text{S}}^{\text{H}}(c_{\text{cr}})$ is up to eight times larger than predicted by $D_{\text{SE}}(c_{\text{cr}})$, highlighting a pronounced decoupling between microscopic and macroscopic transport polymer-crowded environments within the semi-dilute regime.

To elucidate the origin of this behavior, we determined key parameters relevant to protein–polymer solutions. In the semi-dilute regime ($c >c^*$, see Table~\ref{table:overlap}) the correlation length $\xi$ replaces the polymer’s radius of gyration as the characteristic structural length scale governing the depletion layer thickness \cite{Col_RheAct_2010}. $\xi$ defines the distance beyond which polymer–polymer interactions are screened by neighboring chains and was extracted by fitting small angle X-ray scattering (SAXS) curves of pure crowder solutions using an Ornstein-Zernike-like function \cite{Col_RheAct_2010,Ped_JPolymSci_2004,Hig_Oxford_1994}:
\begin{align}
    I(q) = \frac{I(0)}{1 + (q \,\xi)^b},
    \label{eq:O-Z}
\end{align}
where $\xi$ and $b$ are fitting parameters. The extracted correlation lengths are shown in Fig.~\ref{fig:Dc}D. The inset provides example fits of SAXS curves for dextran 100 following Eq.~\ref{eq:O-Z}. 
Since the correlation length defines the depletion layer thickness $\delta_{\text{s}}$ in the semi-dilute regime, this parameter was estimated following the approaches outlined in refs. \cite{Zos_pnas_2020,fleer2003mean,tuinier2002interaction,hanke1999polymer}.
The corresponding equations are provided in the \textit{SI Appendix}.
The resulting $\delta_{\text{s}}$ values as a function of normalized crowder concentration ($c_{\text{cr}}/c^*$) are shown in Fig.~\ref{fig:Dc}E.
Following the approach of Tuinier \textit{et al.}\ \cite{tuinier2006depletion}, the viscosity reduction within the depletion layer, caused by the locally reduced macromolecule concentration and hereafter referred to as the microscopic viscosity ($\eta_{\text{micro}}$), was determined from the measured macroscopic viscosity ($\eta_{\text{macro}}$), the solvent viscosity, and $\delta_{\text{s}}$. Details of this calculation are provided in the \textit{SI Appendix}.
The resulting concentration-dependent ratio $\eta_{\text{macro}}/\eta_{\text{micro}}$ is shown in Fig.~\ref{fig:Dc}F and closely mirrors the trend observed for the self-diffusion constant ratio $D_{\text{S}}^{\text{H}}/D_{\text{SE}}$. 

This analysis indicates that the observed non-monotonic behavior arises from the interplay between microscopic and macroscopic viscosities mediated by depletion effects. Within the depletion layer, where macromolecules are excluded, $\eta_{\text{micro}}$ is significantly reduced. At low concentrations, the large depletion layer surrounding the protein facilitates faster self-diffusion. As the crowder concentration increases, however, the correlation length $\xi$ decreases, leading to a corresponding shrinkage of the depletion layer (Fig. \ref{fig:Dc}D and E). This constrains  protein motion and causes a sharper drop in $D_{\text{S}}^{\text{H}}(c_{\text{cr}})$, which becomes increasingly dominated by the rising bulk viscosity. 

Beyond approximately $\sim 2c^*$, the depletion layer thickness stabilizes, while the macroscopic viscosity continues to increase steeply. As a result, $D_{\text{SE}}(c_{\text{cr}})$ decreases more rapidly, reversing the earlier trend. This crossover near $2c^*$ was consistently observed for all polysaccharide crowders investigated,  suggesting a universal behavior of such systems. 

We also note that the absolute values in Fig.~\ref{fig:Dc}F differ by a factor of two or more from the experimental values in Fig.~\ref{fig:Dc}C, indicating that additional mechanisms beyond the depletion layer must contribute. Enhanced nanoparticle diffusion in macromolecular crowder solutions is well documented and has been attributed to the fact that nanoparticles probe the local polymer structure rather than the bulk properties of the solution \cite{kohli2012diffusion}. We therefore hypothesize that our experiments capture a combination of this established  enhancement effect together with the influence of the nanoscopic depletion layer, giving rise to the observed non-monotonic behavior. We attribute these observations to the high sensitivity of XPCS to molecular-scale dynamics, which enables us to resolve the influence of structural features such as the depletion layer that are less accessible with conventional optical techniques.

\section*{Discussion}

This study demonstrates that different types of crowder molecules induce distinct effects on protein dynamics. For ferritin, sucrose primarily preserves repulsive protein–protein interactions, whereas polysaccharide crowders introduce additional short-range attractions via depletion interactions, a well-established feature of protein–polymer mixtures. These attractions drive the emergence of IRO in ferritin–polysaccharide solutions, a universal effect observed for both dextran and Ficoll. Our results thus provide experimental validation of the simulation results by Riest \textit{et al.}\cite{fries2025chemically}, which predicted that systems with short-range attraction combined with long-range repulsion exhibit unusual collective diffusion and hydrodynamic behavior, characterized by a low-$q$ IRO in $S(q)$.

Despite their distinct conformations in solution -- Ficoll forming compact, highly branched coils and dextran adopting more flexible, extended structures -- short-range attraction consistently emerged once the polysaccharide overlap concentration was exceeded, highlighting the critical role of polymer concentration. Moreover, increasing the molecular weight of dextran enhanced these attractions, producing more stable and longer-lived ferritin complexes stabilized by IRO. This underscores that even variations in molecular weight within the same crowder type can substantially alter crowding effects.

Self-diffusion measurements on both macroscopic and microscopic scales revealed pronounced differences, with the ratio of microscopic to macroscopic diffusion coefficients displaying a non-monotonic dependence on crowder concentration. This behavior arises from the interplay between local depletion effects and bulk viscosity: at low concentrations, the shrinking depletion layer reduces self-diffusion, whereas beyond the semi-dilute regime, the steep increase in macroscopic viscosity becomes dominant. Normalization by the overlap concentration $c^*$ produced a universal scaling behavior, with a turning point around $2c^*$, beyond which macroscopic viscosity overrides depletion-induced mobility enhancements.

Our investigation of ferritin under crowded conditions demonstrate that nanoscale protein dynamics are strongly shaped by: 
i) the interplay between proteins and crowders, including depletion interactions,
ii) crowder type (e.g., small spheres vs.\ polymers) and its concentration as well as concentration-dependent structure, and
iii) the crowder molecular weight. 
As a result, even commonly used crowding agents can induce non-trivial and system-specific behaviors, ephasizing the multifaceted nature of crowding. These insights provide a framework for reconciling the diverse and sometimes contradictory reports on protein behavior under crowded conditions, including enzymatic activity and protein association \cite{jiang2007effects,dhar2010structure,norris2011true,okamoto2010increase,homchaudhuri2006effect,derham2006effect,ren2003effects,wenner1999crowding,pastor2011effect,van1999effects}. 

Ultimately, this work highlights the necessity of integrating polymer physics and depletion theory alongside excluded volume effects into models of crowded environments, as these factors not only govern structural organization but also influence protein dynamics.

\section*{Materials and Methods}
\subsection*{Sample preparation}

Ferritin was purchased from Sigma Aldrich and corresponds to ferritin from equine spleen in saline solution (CAS: 9007-73-2) with an initial protein concentration of $c$ = 71 mg/ml. To increase the protein concentration, 8 ml of this solution was centrifuged at 5000 g and 1.5 ml at 10000 g for 15 minutes using 30 kDa Millipore filters. Using SAXS calibration curves of ferritin solutions of known concentration, the resulting ferritin concentration was determined to be 110 mg/ml and 265 mg/ml. 

Stock solutions of dextran 40, dextran 100 and dextran 500 (Roth, CAS: 9004-54-0) were prepared in concentrations of 20, 40, 43.75 and 50 \%w/w as well as of Ficoll\textsuperscript{\textregistered}400 (Sigma Aldrich, CAS: 26873-85-8) and sucrose (Sigma Aldrich, CAS: 57-50-1) in concentrations of 20, 40, 50 and 60 \%w/w by dissolving them in 103 mM saline solution with a pH value of 7. 

To obtain a final dextran concentration of 35 \%w/w in the sample, the 265 mg/ml ferritin solution was mixed with the 43.75 \%w/w dextran stock in a 20\% : 80\% volume ratio. All the other stock solutions were mixed 50\% : 50\% with the 110 mg/ml ferritin solution. The resulting crowder concentrations in the sample are therefore 10, 20, 25 and 35 \%w/w using dextran as a crowder and 10, 20, 25 and 30 \%w/w using surcrose and Ficoll as crowder. The ferritin concentration remains constant at approx.\ 55 mg/ml. The solutions were filled into quartz capillaries with an outer diameter of 1.5 mm and a wall thickness of 10 $\si{\micro\meter}$. 
\subsection*{Experimental setup}
Two experiments were carried out at the Material Imaging and Dynamics (MID) instrument of European XFEL to collect the data presented. In both cases the small-angle X-ray scattering (SAXS) configuration was used. The first experiment was performed using the pink beam with a photon energy of 9 keV. 200 pulses per train were delivered with a repetition rate of 2.25 MHz, which corresponds to a pulse spacing of 440 ns. For the second experiment the pink beam with a photon energy of 10 keV and a higher repetition rate of 4.5 MHz (220 ns pulse spacing) was available. The inter-train repetition rate was 10 Hz in both experiments. Using compound refractive lenses (CRL), the beam was focused to 14.5 $\si{\micro\meter}$ and 14.2 $\si{\micro\meter}$ in diameter, respectively. The values for the beamsize were obtained by fitting the speckle contrast in conjunction with an effective X-ray bandwidth of $\Delta E/E$ = $13 \times 10^{-3}$ and $9\times 10^{-3}$, respectively. Using a scanning speed of 0.4 mm/s of the Fast Solid Sample Scanner (FSSS) ensures the refreshment of the sample volume in between the trains, as described in ref. \cite{Reiser_natcomm_2022}. The data acquisition was realized by the AGIPD \cite{All_synrad_2019,Szt_frontphys_2024}, which was positioned 7.15 m and 7.68 m behind the sample. Thus, a $q$-range of 0.1 to 1 nm$^{-1}$ and 0.075 to 1 nm$^{-1}$ was accessible, respectively. All measurements were performed at room temperature (23°C). 
\subsection*{XPCS calculations}
The scattering of coherent X-rays by the ferritin particles within the sample system generates a speckle pattern that varies over time as a result of particle motion. To analyze the dynamics of ferritin, the fluctuating intensities measured experimentally at two distinct times, $t_1$ and $t_2$, were correlated. This approach employs two-time correlation functions of the form \cite{Lehmxpcstheory,Perxpcstheory,sutton2003using}:
\begin{equation}
    C(q, t_1, t_2) = \frac{\langle I(q,t_1)I(q,t_2)\rangle_{\mathrm{pix}}}{\langle I(q,t_1)\rangle_{\mathrm{pix}}\langle I(q,t_2)\rangle_{\mathrm{pix}}}
\end{equation}
where $\langle$...$\rangle_{\mathrm{pix}}$ indicate the average of all detector pixels within the same $q$-range. In the first experiment, the $q$-binning was set to $\Delta q$ = 0.045 nm$^{-1}$, in the second to $\Delta q$ = 0.04 nm$^{-1}$. The calculations were perfomed utilizing the EXtra-speckle data processing pipeline \cite{leonoau2025pipeline} coupled to DAMNIT tool \cite{damnit} developed at EuXFEL. For a sufficient signal-to-noise ratio, data were collected and averaged over several thousand trains. 
The intensity autocorrelation function, $g_2$($q$,$t$) = $\langle C(q,t_1, t_1+t) \rangle _{t_1}$, were extracted by performing horizontal cuts along the $t_1$ axis, starting from the diagonal \cite{bikondoa2017use}. Based on the standard deviation of the fluctuating contrast values of the TTC, a weighted average over trains and times was calculated to determine the final $g_2$($q$,$t$) function. The error bars correspond to the square root of the inverse of the summed weights. 

\section*{Acknowledgements}
We acknowledge European XFEL in Schenefeld, Germany, for providing X-ray free-electron laser beamtime at the Materials Imaging and Dynamics (MID) instrument and would like to thank the staff for their support. The data presented here were collected during two beamtimes allocated to proposals 3348 and 5397. This work was supported by the Maxwell computational resources of the Deutsches Elektronen-Synchrotron DESY, Hamburg, Germany. Furthermore, we acknowledge DELTA machine group for providing synchrotron radiation and for technical support. We thank Marvin Kowalski, Jenny Schrage, Pia Kwiasowski, Maximilian Senft, Prince Prabhu Rajaiah, Lara Reichart, and Chang Hee Woo for their collaboration during beamtime and Randeer Gautam for the discussion of the experimental results. 

\paragraph*{Funding:}
We acknowledge BMBF (05K19PS1, 05K20PSA and 05K22PS1)(to C.G.); 05K19V\linebreak TB, (to F.S. and F.Z.), DFG-ANR (SCHR700/28-1, SCHR700/42-1, ANR-16-CE92-0009) (to F.S. and F.Z.). We acknowledge funding from NFDI 40/1 (DAPHNE4NFDI). F.P., I.A.A., and A.G. acknowledge funding from the European Union’s Horizon Europe research and innovation programme under the Marie Skłodowska-Curie grant agreement No. 101081419 (PRISMAS) (F.P. and I.A.A.) and 101149230 (CRYSTAL-X) (F.P. and A.G.). F.P. acknowledges financial support by the Swedish National Research Council (Vetenskapsrådet) under Grant No. 2019-05542, 2023-05339 and within the Röntgen-Ångström Cluster Grant No. 2019-06075, and the kind financial support from Knut och Alice Wallenberg foundation (WAF, Grant. No. 2023.0052). This research is supported by the Center of Molecular Water Science (CMWS) of DESY in an Early Science Project, the MaxWater initiative of the Max-Planck-Gesellschaft (Project No. CTS21:1589), Carl Tryggers and the Wenner-Gren Foundations (Project No. UPD2021-0144). 

\section*{Author contributions} 
M.D. analyzed data. N.D.A., M.P., C.G., F.Z., F.S., S.R., A.G., F.P., J.M., and F.L. discussed data. M.D., N.D.A., A.G., M.B., and J.S. prepared the samples. M.D., N.D.A., A.G., S.R., S.T., A.M.R., A.L., F.U., I.A., M.B., F.L., M.P., and C.G. conducted the experiment. J.M., W.J., F.B., U.B., J.H., J.P., A.R., R.S., M.Y., A.Z., and A.M. operated MID instrument. J.M., F.B., and A.L., provided the data analysis pipeline. M.D., M.P., and C.G. wrote the manuscript with all authors' input.
\paragraph*{Competing interests}
The authors declare no competing interest.

\paragraph*{Data availability}
Data recorded at the European XFEL are available at DOI: 10.22003/XFEL.\linebreak EU-DATA-003348-00 and DOI:10.22003/XFEL.EU-DATA-005397-00. The data collected at DELTA and the code used to analyze the data is available from the corresponding authors upon request.
\newpage
\bibliographystyle{pnas-new}
\bibliography{main}

\begin{thebibliography}{10}

\bibitem{Gra_JCollScie_1958}
Granath KA (1958) Solution properties of branched dextrans.
\newblock {\em J. Colloid Sci.} 13(4):308--328.

\bibitem{ranganathan2022ficoll_2}
Ranganathan VT, Bazmi S, Wallin S, Liu Y, Yethiraj A (2022) Is {Ficoll} a colloid or polymer? {A} multitechnique study of a prototypical excluded-volume macromolecular crowder.
\newblock {\em Macromolecules} 55(20):9103--9112.

\bibitem{spencer2024concentration}
Spencer SJ, Ranganathan VT, Yethiraj A, Andrews GT (2024) Concentration dependence of elastic and viscoelastic properties of aqueous solutions of {Ficoll} and bovine serum albumin by brillouin light scattering spectroscopy.
\newblock {\em Langmuir} 40(9):4615--4622.

\bibitem{Bie_PLoS_2019}
Biehl R (2019) Jscatter, a program for evaluation and analysis of experimental data.
\newblock {\em PLoS ONE} 14(6).

\bibitem{Liu_ChemPhys_2005}
Liu Y, Chen WR, Chen SH (2005) Cluster formation in {two-Yukawa} fluids.
\newblock {\em J. Chem. Phys.} 122(4).

\bibitem{Wie_JApplCryst_2021_2}
Wieland DCF, et~al. (2021) {ASAXS} measurements on ferritin and apoferritin at the {bioSAXS} beamline p12 {(PETRA III, DESY)}.
\newblock {\em J. Appl. Cryst.} 54:830--838.

\bibitem{Bee_PhysAStat_1983_2}
Beenakker C, Mazur P (1983) Self-diffusion of spheres in a concentrated suspension.
\newblock {\em Phys. A Stat. Mech. Its Appl.} 120:388--410.

\bibitem{Bee_PhysAStat_1984_2}
Beenakker C, Mazur P (1984) Diffusion of spheres in a concentrated suspension {II}.
\newblock {\em Phys. A Stat. Mech. Its Appl.} 126:349--370.

\bibitem{Gen_PhysAStat_1991}
Genz U, Klein R (1991) Collective diffusion of charged spheres in the presence of hydrodynamic interaction.
\newblock {\em Phys. A Stat. Mech. Its Appl.} 171:26--42.

\bibitem{menon1991new}
Menon S, Manohar C, Rao KS (1991) A new interpretation of the sticky hard sphere model.
\newblock {\em J. Chem. Phys.} 95(12):9186--9190.

\bibitem{sztucki2006kinetic}
Sztucki M, et~al. (2006) Kinetic arrest and glass-glass transition in short-ranged attractive colloids.
\newblock {\em Phys. Rev. E.} 74(5):051504.

\bibitem{Zos_pnas_2020_2}
Zosel F, Soranno A, Buholzer KJ, Nettels D, Schuler B (2020) Depletion interactions modulate the binding between disordered proteins in crowded environments.
\newblock {\em Proc. Natl. Acad. Sci. U.S.A.} 117(24):13480--13489.

\bibitem{fleer2003mean_2}
Fleer GJ, Skvortsov AM, Tuinier R (2003) Mean-field equation for the depletion thickness.
\newblock {\em Macromolecules} 36(20):7857--7872.

\bibitem{tuinier2002interaction_2}
Tuinier R, Lekkerkerker H, Aarts D (2002) Interaction potential between two spheres mediated by excluded volume polymers.
\newblock {\em Phys. Rev. E} 65(6):060801.

\bibitem{hanke1999polymer_2}
Hanke A, Eisenriegler E, Dietrich S (1999) Polymer depletion effects near mesoscopic particles.
\newblock {\em Phys. Rev. E} 59(6):6853.

\bibitem{tuinier2006depletion_2}
Tuinier R, Dhont J, Fan TH (2006) How depletion affects sphere motion through solutions containing macromolecules.
\newblock {\em Europhys. Lett.} 75(6):929.

\end{thebibliography}


\begin{thebibliography}{100}

\bibitem{ellis2001macromolecular}
Ellis RJ (2001) Macromolecular crowding: obvious but underappreciated.
\newblock {\em Trends Biochem. Sci.} 26(10):597--604.

\bibitem{laurent1963interaction}
Laurent TC, Ogston A (1963) The interaction between polysaccharides and other macromolecules. 4. {The} osmotic pressure of mixtures of serum albumin and hyaluronic acid.
\newblock {\em Biochemical Journal} 89(2):249.

\bibitem{minton1981excluded}
Minton AP (1981) Excluded volume as a determinant of macromolecular structure and reactivity.
\newblock {\em Biopolymers} 20(10):2093--2120.

\bibitem{minton1997influence}
Minton AP (1997) Influence of excluded volume upon macromolecular structure and associations in "crowded” media.
\newblock {\em Curr. Opin. Biotechnol.} 8(1):65--69.

\bibitem{grimaldo2019dynamics}
Grimaldo M, Roosen-Runge F, Zhang F, Schreiber F, Seydel T (2019) Dynamics of proteins in solution.
\newblock {\em Q. Rev. Biophys.} 52:e7.

\bibitem{yu2016biomolecular}
Yu I, et~al. (2016) Biomolecular interactions modulate macromolecular structure and dynamics in atomistic model of a bacterial cytoplasm.
\newblock {\em elife} 5:e19274.

\bibitem{roosen2011protein}
Roosen-Runge F, et~al. (2011) Protein self-diffusion in crowded solutions.
\newblock {\em Proc. Natl. Acad. Sci. U. S. A.} 108(29):11815--11820.

\bibitem{luby1999cytoarchitecture}
Luby-Phelps K (1999) Cytoarchitecture and physical properties of cytoplasm: volume, viscosity, diffusion, intracellular surface area.
\newblock {\em Int. Rev. Cytol.} 192:189--221.

\bibitem{heinen2012viscosity}
Heinen M, et~al. (2012) Viscosity and diffusion: crowding and salt effects in protein solutions.
\newblock {\em Soft Matter} 8(5):1404--1419.

\bibitem{dix2008crowding}
Dix JA, Verkman A (2008) Crowding effects on diffusion in solutions and cells.
\newblock {\em Annu. Rev. Biophys.} 37(1):247--263.

\bibitem{gnutt2016macromolecular}
Gnutt D, Ebbinghaus S (2016) The macromolecular crowding effect - from in vitro into the cell.
\newblock {\em Biol. Chem.} 397(1):37--44.

\bibitem{homchaudhuri2006effect}
Homchaudhuri L, Sarma N, Swaminathan R (2006) Effect of crowding by dextrans and {Ficolls} on the rate of alkaline phosphatase--catalyzed hydrolysis: {A} size-dependent investigation.
\newblock {\em Biopolymers} 83(5):477--486.

\bibitem{derham2006effect}
Derham BK, Harding JJ (2006) The effect of the presence of globular proteins and elongated polymers on enzyme activity.
\newblock {\em Biochim. Biophys. Acta - Proteins Proteom.} 1764(6):1000--1006.

\bibitem{ren2003effects}
Ren G, Lin Z, Tsou Cl, Wang Cc (2003) Effects of macromolecular crowding on the unfolding and the refolding of d-glyceraldehyde-3-phosophospate dehydrogenase.
\newblock {\em J. Protein Chem.} 22:431--439.

\bibitem{wenner1999crowding}
Wenner JR, Bloomfield VA (1999) Crowding effects on {EcoRV} kinetics and binding.
\newblock {\em Biophys. J.} 77(6):3234--3241.

\bibitem{ranganathan2022ficoll}
Ranganathan VT, Bazmi S, Wallin S, Liu Y, Yethiraj A (2022) Is {Ficoll} a colloid or polymer? {A} multitechnique study of a prototypical excluded-volume macromolecular crowder.
\newblock {\em Macromolecules} 55(20):9103--9112.

\bibitem{biswas2018mixed}
Biswas S, Kundu J, Mukherjee SK, Chowdhury PK (2018) Mixed macromolecular crowding: {A} protein and solvent perspective.
\newblock {\em ACS omega} 3(4):4316--4330.

\bibitem{rusinga2017soft}
Rusinga FI, Weis DD (2017) Soft interactions and volume exclusion by polymeric crowders can stabilize or destabilize transient structure in disordered proteins depending on polymer concentration.
\newblock {\em Proteins: Struct., Funct., Bioinf.} 85(8):1468--1479.

\bibitem{colby2010structure}
Colby RH (2010) Structure and linear viscoelasticity of flexible polymer solutions: comparison of polyelectrolyte and neutral polymer solutions.
\newblock {\em Rheol. acta} 49:425--442.

\bibitem{teraoka2002polymer}
Teraoka I (2002) {\em Polymer solutions}.
\newblock (Wiley).

\bibitem{hirata2003small}
Hirata Y, et~al. (2003) Small-angle {X}-ray scattering studies of moderately concentrated dextran solution.
\newblock {\em Carbohydr. Polym.} 53(3):331--335.

\bibitem{höfling2013anomalous}
Höfling F, Franosch T (2013) Anomalous transport in the crowded world of biological cells.
\newblock {\em Rep. Prog. Phys.} 76(4):046602.

\bibitem{metzler2014anomalous}
Metzler R, Jeon JH, Cherstvy AG, Barkai E (2014) Anomalous diffusion models and their properties: non-stationarity{,} non-ergodicity{,} and ageing at the centenary of single particle tracking.
\newblock {\em Phys. Chem. Chem. Phys.} 16(44):24128--24164.

\bibitem{srinivasan2024breaking}
Srinivasan H, Sharma VK, Mitra S (2024) Breaking the {B}rownian barrier: models and manifestations of molecular diffusion in complex fluids.
\newblock {\em Phys. Chem. Chem. Phys.} 26(47):29227--29250.

\bibitem{szymanski2009elucidating}
Szymanski J, Weiss M (2009) Elucidating the origin of anomalous diffusion in crowded fluids.
\newblock {\em Phys. Rev. Lett.} 103(3):038102.

\bibitem{sabri2020elucidating}
Sabri A, Xu X, Krapf D, Weiss M (2020) Elucidating the origin of heterogeneous anomalous diffusion in the cytoplasm of mammalian cells.
\newblock {\em Phys. Rev. Lett.} 125(5):058101.

\bibitem{asakura1954interaction}
Asakura S, Oosawa Fm (1954) On interaction between two bodies immersed in a solution of macromolecules.
\newblock {\em J. Chem. Phys.} 22(7):1255--1256.

\bibitem{vrij1976polymers}
Vrij A (1976) Polymers at interfaces and the interactions in colloidal dispersions.
\newblock {\em Pure Appl. Chem.} 48(4):471--483.

\bibitem{tuinier2011colloids}
Tuinier R, Lekkerkerker HN (2011) {\em Colloids and the depletion interaction}.
\newblock (Springer Netherlands).

\bibitem{roth2010fundamental}
Roth R (2010) Fundamental measure theory for hard-sphere mixtures: a review.
\newblock {\em J. Phys.: Condens. Matter} 22(6):063102.

\bibitem{roth2000depletion}
Roth R, Evans R, Dietrich S (2000) Depletion potential in hard-sphere mixtures: Theory and applications.
\newblock {\em Phys. Rev. E} 62(4):5360.

\bibitem{bechinger1999understanding}
Bechinger C, Rudhardt D, Leiderer P, Roth R, Dietrich S (1999) Understanding depletion forces beyond entropy.
\newblock {\em Phys. Rev. Lett.} 83(19):3960.

\bibitem{neu2006depletion}
Neu B, Meiselman HJ (2006) Depletion interactions in polymer solutions promote red blood cell adhesion to albumin-coated surfaces.
\newblock {\em Biochim. Biophys. Acta - Gen. Subj.} 1760(12):1772--1779.

\bibitem{smith1995depletion}
Smith NJ, Williams PA (1995) Depletion flocculation of polystyrene latices by water-soluble polymers.
\newblock {\em J. Chem. Soc. Faraday Trans.} 91(10):1483--1489.

\bibitem{zhang2023depletion}
Zhang H, Kong D, Zhang W, Liu H (2023) Depletion attraction in colloidal and bacterial systems.
\newblock {\em Front. Mater.} 10:1206819.

\bibitem{gogelein2009polymer}
G{\"o}gelein C, N{\"a}gele G, Buitenhuis J, Tuinier R, Dhont JK (2009) Polymer depletion-driven cluster aggregation and initial phase separation in charged nanosized colloids.
\newblock {\em J. Chem. Phys.} 130(20).

\bibitem{tuinier2000depletion}
Tuinier R, Dhont J, De~Kruif C (2000) Depletion-induced phase separation of aggregated whey protein colloids by an exocellular polysaccharide.
\newblock {\em Langmuir} 16(4):1497--1507.

\bibitem{von2019dynamic}
von B{\"u}low S, Siggel M, Linke M, Hummer G (2019) Dynamic cluster formation determines viscosity and diffusion in dense protein solutions.
\newblock {\em Proc. Natl. Acad. Sci. U.S.A.} 116(20):9843--9852.

\bibitem{liu2011lysozyme}
Liu Y, et~al. (2011) Lysozyme protein solution with an intermediate range order structure.
\newblock {\em J. Phys. Chem. B} 115(22):7238--7247.

\bibitem{piazza2006micro}
Piazza R, et~al. (2006) Micro-heterogeneity and aggregation in $\beta$ 2-microglobulin solutions: effects of temperature, {pH}, and conformational variant addition.
\newblock {\em Eur. Biophys. J.} 35:439--445.

\bibitem{kovalchuk2009formation}
Kovalchuk N, Starov V, Langston P, Hilal N (2009) Formation of stable clusters in colloidal suspensions.
\newblock {\em Adv. Colloid Interface Sci.} 147:144--154.

\bibitem{godfrin2014generalized}
Godfrin PD, Valadez-P{\'e}rez NE, Castaneda-Priego R, Wagner NJ, Liu Y (2014) Generalized phase behavior of cluster formation in colloidal dispersions with competing interactions.
\newblock {\em Soft {M}atter} 10(28):5061--5071.

\bibitem{riest2015short}
Riest J, N{\"a}gele G (2015) Short-time dynamics in dispersions with competing short-range attraction and long-range repulsion.
\newblock {\em Soft Matter} 11(48):9273--9280.

\bibitem{Reiser_natcomm_2022}
Reiser M, et~al. (2022) Resolving molecular diffusion and aggregation of antibody proteins with megahertz {X-ray} free-electron laser pulses.
\newblock {\em Nat. Commun.} 13(5528).

\bibitem{Gir_arx_2024}
Girelli A, et~al. (2025) Coherent {X-rays} reveal anomalous molecular diffusion and cage effects in crowded protein solutions.
\newblock {\em Nat. Commun.}
\newblock In print. Preprint available at \url{https://arxiv.org/abs/2410.08873}.

\bibitem{kohli2012diffusion}
Kohli I, Mukhopadhyay A (2012) Diffusion of nanoparticles in semidilute polymer solutions: {Effect} of different length scales.
\newblock {\em Macromolecules} 45(15):6143--6149.

\bibitem{Theil_wiley_1990}
Theil EC (1990) {\em The Ferritin Family of Iron Storage Proteins}.
\newblock (John Wiley \& Sons, Ltd), pp. 421--449.

\bibitem{Har_bba_1996}
Harrison PM, Arosio P (1996) The ferritins: molecular properties, iron storage function and cellular regulation.
\newblock {\em Biochim. Biophys. Acta - Bioenerg.} 1275(3):161--203.

\bibitem{Cha_structuralbio_1999}
Chasteen ND, Harrison PM (1999) The ferritins: molecular properties, iron storage function and cellular regulation.
\newblock {\em J. Struct. Biol.} 126:182–194.

\bibitem{Moh_biomacrom_2021}
Mohanty A, K M, Jena SS, Behera RK (2021) Kinetics of ferritin self-assembly by laser light scattering: Impact of subunit concentration, {pH}, and ionic strength.
\newblock {\em Biomacromolecules} 22:1389--1398.

\bibitem{Rod_pharma_2021}
Rodrigues MQ, Alves PM, Roldão A (2021) Functionalizing ferritin nanoparticles for vaccine development.
\newblock {\em Pharmaceutics} 13(1621).

\bibitem{Khos_contrrelease_2018}
Khoshnejad M, Parhiz H, Shuvaev VV, Dmochowski IJ, Muzykantov VR (2018) Ferritin-based drug delivery systems: Hybrid nanocarriers for vascular immunotargeting.
\newblock {\em J. Control. Release} 282:13--24.

\bibitem{Luc_Interbiomacrom_2022}
Lucignano R, et~al. (2024) Human ferritin nanocarriers for drug-delivery: A molecular view of the disassembly process.
\newblock {\em Int. J. Biol. Macromol.} 277(2).

\bibitem{madsen2021materials}
Madsen A, et~al. (2021) Materials {Imaging and Dynamics} ({MID}) instrument at the {European X-ray Free-Electron Laser Facility}.
\newblock {\em J. Synchrotron Radiat.} 28(2):637--649.

\bibitem{anthuparambil2024salt}
Anthuparambil ND, et~al. (2024) Salt induced slowdown of kinetics and dynamics during thermal gelation of egg-yolk.
\newblock {\em J. Chem. Phys.} 161(5).

\bibitem{anthuparambil2023exploring}
Anthuparambil ND, et~al. (2023) Exploring non-equilibrium processes and spatio-temporal scaling laws in heated egg yolk using coherent {X-rays}.
\newblock {\em Nat. Commun.} 14(1):5580.

\bibitem{timmermann2023x}
Timmermann S, et~al. (2023) X-ray driven and intrinsic dynamics in protein gels.
\newblock {\em Sci. Rep.} 13(1):11048.

\bibitem{girelli2021microscopic}
Girelli A, et~al. (2021) Microscopic dynamics of liquid-liquid phase separation and domain coarsening in a protein solution revealed by {X-ray} photon correlation spectroscopy.
\newblock {\em Phys. Rev. Lett.} 126(13):138004.

\bibitem{begam2021kinetics}
Begam N, et~al. (2021) Kinetics of network formation and heterogeneous dynamics of an egg white gel revealed by coherent {X-ray} scattering.
\newblock {\em Phys. Rev. Lett.} 126(9):098001.

\bibitem{dallari2021microsecond}
Dallari F, et~al. (2021) Microsecond hydrodynamic interactions in dense colloidal dispersions probed at the {European XFEL}.
\newblock {\em IUCrJ} 8(5):775--783.

\bibitem{anthuparambil2025softness}
Anthuparambil ND, et~al. (2025) Softness and hydrodynamic interactions regulate lipoprotein transport in crowded yolk environments.
\newblock {\em arXiv preprint arXiv:2505.22520}.

\bibitem{dargasz2022x}
Dargasz M, et~al. (2022) X-ray scattering at beamline {BL2 of DELTA: Studies} of lysozyme-lysozyme interaction in heavy water and structure formation in 1-hexanol in {\em J. Phys.: Conf. Ser.}
\newblock Vol.{} 2380, p. 012031.

\bibitem{Willxpcstheory}
Williams G, Watts DCN (1970) Non-symmetrical dielectric relaxation behaviour arising from a simple empirical decay function.
\newblock {\em Trans. Faraday Soc.} 66(80).

\bibitem{Hru_PhysRevLett_2012}
Hruszkewycz S, et~al. (2012) High contrast {X}-ray speckle from atomic-scale order in liquids and glasses.
\newblock {\em Phys. Rev. Lett.} 109:185502.

\bibitem{Leh_pnas_2020}
Lehmkühler F, et~al. (2020) Emergence of anomalous dynamics in soft matter probed at the {European} {XFEL}.
\newblock {\em Proc. Natl. Acad. Sci. U.S.A.} 117:24110–24116.

\bibitem{omari2009diffusion}
Omari RA, Aneese AM, Grabowski CA, Mukhopadhyay A (2009) Diffusion of nanoparticles in semidilute and entangled polymer solutions.
\newblock {\em J. Phys. Chem. B} 113(25):8449--8452.

\bibitem{banks2005anomalous}
Banks DS, Fradin C (2005) Anomalous diffusion of proteins due to molecular crowding.
\newblock {\em Biophys. J.} 89(5):2960--2971.

\bibitem{sanabria2007multiple}
Sanabria H, Kubota Y, Waxham MN (2007) Multiple diffusion mechanisms due to nanostructuring in crowded environments.
\newblock {\em Biophys. J.} 92(1):313--322.

\bibitem{mazurkiewicz1998viscosity}
Mazurkiewicz J, Nowotny-R{\'o}{\.z}a{\'n}ska M (1998) Viscosity of aqueous solutions of saccharides.
\newblock {\em Pol. J. Food Nutr. Sci.} 48:171--180.

\bibitem{masuelli2014dextrans}
Masuelli MA (2013) Dextrans in aqueous solution. {Experimental} review on intrinsic viscosity measurements and temperature effect.
\newblock {\em J. Polym. Biopolym. Phys. Chem.} 1:13--21.

\bibitem{Val_JInorgBiochem_2012}
de~Val N, Declercq JP, Limc CK, Crichton RR (2012) Structural analysis of haemin demetallation by {L-chain} apoferritins.
\newblock {\em J. Inorg. Biochem.} 112:77--84.

\bibitem{Lehmxpcstheory}
Lehmkühler F, Roseker W, Grübel G (2021) From femtoseconds to hours - {Measuring} dynamics over 18 orders of magnitude with coherent {X-rays}.
\newblock {\em Appl. Sci.} 11(6179).

\bibitem{zhang2024effective}
Zhang F, et~al. (2024) Effective interactions in protein solutions with and without clustering.
\newblock {\em Phys. A: Stat. Mech. Appl.} 650:129995.

\bibitem{Por_PhysChemLett_2010}
Porcar L, et~al. (2010) Formation of the dynamic clusters in concentrated lysozyme protein solutions.
\newblock {\em J. Phys. Chem. Lett.} 1(1):126--129.

\bibitem{Liu_PhysChemB_2011}
Liu Y, et~al. (2011) Lysozyme protein solution with an intermediate range order structure.
\newblock {\em J. Phys. Chem. B} 115(22):7238--7247.

\bibitem{Yea_biophys_2014}
Yearley EJ, et~al. (2014) Observation of small cluster formation in concentrated monoclonal antibody solutions and its implications to solution viscosity.
\newblock {\em Biophys. J.} 106:1763--1770.

\bibitem{semenov2008theory}
Semenov A (2008) Theory of colloid stabilization in semidilute polymer solutions.
\newblock {\em Macromolecules} 41(6):2243--2249.

\bibitem{shvets2013effective}
Shvets A, Semenov A (2013) Effective interactions between solid particles mediated by free polymer in solution.
\newblock {\em J. Chem. Phys.} 139(5).

\bibitem{semenov2015theory}
Semenov A, Shvets A (2015) Theory of colloid depletion stabilization by unattached and adsorbed polymers.
\newblock {\em Soft Matter} 11(45):8863--8878.

\bibitem{gogelein2008phase}
G{\"o}gelein C, Tuinier R (2008) Phase behaviour of a dispersion of charge-stabilised colloidal spheres with added non-adsorbing interacting polymer chains.
\newblock {\em Eur. Phys. J. E} 27:171--184.

\bibitem{nagele1996dynamics}
N{\"a}gele G (1996) On the dynamics and structure of charge-stabilized suspensions.
\newblock {\em Phys. Rep.} 272(5-6):215--372.

\bibitem{Bee_PhysAStat_1984}
Beenakker C, Mazur P (1984) Diffusion of spheres in a concentrated suspension {II}.
\newblock {\em Phys. A Stat. Mech. Its Appl.} 126:349--370.

\bibitem{Bee_PhysAStat_1983}
Beenakker C, Mazur P (1983) Self-diffusion of spheres in a concentrated suspension.
\newblock {\em Phys. A Stat. Mech. Its Appl.} 120:388--410.

\bibitem{li2011analysis}
Li J, Busscher HJ, Norde W, Sjollema J (2011) Analysis of the contribution of sedimentation to bacterial mass transport in a parallel plate flow chamber.
\newblock {\em Colloids Surf. B: Biointerfaces} 84(1):76--81.

\bibitem{hinderliter2010isdd}
Hinderliter PM, et~al. (2010) {ISDD}: A computational model of particle sedimentation, diffusion and target cell dosimetry for in vitro toxicity studies.
\newblock {\em Part. Fibre Toxicol.} 7:1--20.

\bibitem{ilett1995phase}
Ilett SM, Orrock A, Poon W, Pusey P (1995) Phase behavior of a model colloid-polymer mixture.
\newblock {\em Phys. Rev. E} 51(2):1344.

\bibitem{chatterjee1998microscopic}
Chatterjee AP, Schweizer KS (1998) Microscopic theory of polymer-mediated interactions between spherical particles.
\newblock {\em J. Chem. Phys.} 109(23):10464--10476.

\bibitem{Car_PhysChemB_2011}
Cardinaux F, et~al. (2011) Cluster-driven dynamical arrest in concentrated lysozyme solutions.
\newblock {\em J. Phys. Chem. B} 115(22):7227--7237.

\bibitem{abkenar2017dissociation}
Abkenar M, Gray TH, Zaccone A (2017) Dissociation rates from single-molecule pulling experiments under large thermal fluctuations or large applied force.
\newblock {\em Phys. Rev. E.} 95(4):042413.

\bibitem{Wie_JApplCryst_2021}
Wieland DCF, et~al. (2021) {ASAXS} measurements on ferritin and apoferritin at the {bioSAXS} beamline p12 {(PETRA III, DESY)}.
\newblock {\em J. Appl. Cryst.} 54:830--838.

\bibitem{berezhkovskii2016theory}
Berezhkovskii AM, Szabo A (2016) Theory of crowding effects on bimolecular reaction rates.
\newblock {\em J. Phys. Chem. B} 120(26):5998--6002.

\bibitem{Zos_pnas_2020}
Zosel F, Soranno A, Buholzer KJ, Nettels D, Schuler B (2020) Depletion interactions modulate the binding between disordered proteins in crowded environments.
\newblock {\em Proc. Natl. Acad. Sci. U.S.A.} 117(24):13480--13489.

\bibitem{zhou2008macromolecular}
Zhou HX, Rivas G, Minton AP (2008) Macromolecular crowding and confinement: biochemical, biophysical, and potential physiological consequences.
\newblock {\em Annu. Rev. Biophys.} 37(1):375--397.

\bibitem{schreiber2009fundamental}
Schreiber G, Haran G, Zhou HX (2009) Fundamental aspects of protein-protein association kinetics.
\newblock {\em Chem. Rev.} 109(3):839--860.

\bibitem{kuznetsova2014macromolecular}
Kuznetsova IM, Turoverov KK, Uversky VN (2014) What macromolecular crowding can do to a protein.
\newblock {\em Int. J. Mol. Sci.} 15(12):23090--23140.

\bibitem{mittal2015macromolecular}
Mittal S, Chowhan RK, Singh LR (2015) Macromolecular crowding: Macromolecules friend or foe.
\newblock {\em Biochim. Biophys. Acta - Gen. Subj.} 1850(9):1822--1831.

\bibitem{Col_RheAct_2010}
Colby R (2010) Structure and linear viscoelasticity of flexible polymer solutions: comparison of polyelectrolyte and neutral polymer solutions.
\newblock {\em Rheol. Acta} 49:425--442.

\bibitem{Ped_JPolymSci_2004}
Pedersen JS, Schurtenberger P (2004) Scattering functions of semidilute solutions of polymers in a good solvent.
\newblock {\em J. Polym. Sci. B Polym. Phys.} 42(17):3081--3094.

\bibitem{Hig_Oxford_1994}
Higgins JS, Benoit HC (1994) Polymers and neutron scattering.
\newblock {\em Oxf. Univ. Press}.

\bibitem{fleer2003mean}
Fleer GJ, Skvortsov AM, Tuinier R (2003) Mean-field equation for the depletion thickness.
\newblock {\em Macromolecules} 36(20):7857--7872.

\bibitem{tuinier2002interaction}
Tuinier R, Lekkerkerker H, Aarts D (2002) Interaction potential between two spheres mediated by excluded volume polymers.
\newblock {\em Phys. Rev. E} 65(6):060801.

\bibitem{hanke1999polymer}
Hanke A, Eisenriegler E, Dietrich S (1999) Polymer depletion effects near mesoscopic particles.
\newblock {\em Phys. Rev. E} 59(6):6853.

\bibitem{tuinier2006depletion}
Tuinier R, Dhont J, Fan TH (2006) How depletion affects sphere motion through solutions containing macromolecules.
\newblock {\em Europhys. Lett.} 75(6):929.

\bibitem{fries2025chemically}
Fries J, et~al. (2025) Chemically active droplets in crowded environments.
\newblock {\em arXiv preprint arXiv:2505.11188}.

\bibitem{jiang2007effects}
Jiang M, Guo Z (2007) Effects of macromolecular crowding on the intrinsic catalytic efficiency and structure of enterobactin-specific isochorismate synthase.
\newblock {\em J. Am. Chem. Soc.} 129(4):730--731.

\bibitem{dhar2010structure}
Dhar A, et~al. (2010) Structure, function, and folding of phosphoglycerate kinase are strongly perturbed by macromolecular crowding.
\newblock {\em Proc. Natl. Acad. Sci. U. S. A.} 107(41):17586--17591.

\bibitem{norris2011true}
Norris MG, Malys N (2011) What is the true enzyme kinetics in the biological system? {An} investigation of macromolecular crowding effect upon enzyme kinetics of glucose-6-phosphate dehydrogenase.
\newblock {\em Biochem. Biophys. Res. Commun.} 405(3):388--392.

\bibitem{okamoto2010increase}
Okamoto DN, et~al. (2010) Increase of {SARS-CoV 3CL} peptidase activity due to macromolecular crowding effects in the milieu composition.
\newblock {\em Biol. Chem.} 391:1461--1468.

\bibitem{pastor2011effect}
Pastor I, et~al. (2011) Effect of crowding by dextrans on the hydrolysis of {N-Succinyl-l-phenyl-Ala-p-nitroanilide} catalyzed by $\alpha$-chymotrypsin.
\newblock {\em J. Phys. Chem. B} 115(5):1115--1121.

\bibitem{van1999effects}
Van Den~Berg B, Ellis RJ, Dobson CM (1999) Effects of macromolecular crowding on protein folding and aggregation.
\newblock {\em EMBO J.}

\bibitem{All_synrad_2019}
Allahgholi A, et~al. (2019) The {Adaptive Gain Integrating Pixel Detector} at the {European XFEL}.
\newblock {\em J. Synchrotron Radiat.} 26:74--82.

\bibitem{Szt_frontphys_2024}
Sztuk-Dambietz J, et~al. (2024) Operational experience with {Adaptive Gain Integrating Pixel Detectors} at {European XFEL}.
\newblock {\em Front. Phys.} 11.

\bibitem{Perxpcstheory}
Perakis F, Gutt C (2020) Towards molecular movies with {X-ray} photon correlation spectroscopy.
\newblock {\em Phys. Chem. Chem. Phys.} 22:19443–19453.

\bibitem{sutton2003using}
Sutton M, Laaziri K, Livet F, Bley F (2003) Using coherence to measure two-time correlation functions.
\newblock {\em Opt. Express} 11(19):2268--2277.

\bibitem{leonoau2025pipeline}
Leonau A, et~al. (2025) A pipeline for megahertz {X-ray} photon correlation spectroscopy on soft matter samples at the {MID} instrument of {European XFEL}.
\newblock {\em arXiv preprint arXiv:2506.08668}.

\bibitem{damnit}
{Data Analysis group at the European XFEL} (2025) Damnit user documentation (\url{https://damnit.readthedocs.io/en/latest/}).
\newblock Accessed: 2025-09-02.

\bibitem{bikondoa2017use}
Bikondoa O (2017) On the use of two-time correlation functions for {X-ray} photon correlation spectroscopy data analysis.
\newblock {\em J. Appl. Crystallogr.} 50(2):357--368.

\end{thebibliography}
\newpage 
\setcounter{page}{1}
\title{\huge Supporting Information: \\\LARGE Depletion-Induced Interactions Modulate Nanoscale Protein Diffusion in Polymeric Crowder Solutions \\}
\date{%
  \small
  $^{1}$Department Physik, Universität Siegen, Walter-Flex-Strasse 3, 57072 Siegen, Germany\\[0.5ex]
  $^{2}$Institut für Angewandte Physik, Universität Tübingen, Auf der Morgenstelle 10, 72076 Tübingen, Germany\\[0.5ex]
  $^{3}$Department of Physics, AlbaNova University Center, Stockholm University, 10691 Stockholm, Sweden\\[0.5ex]
  $^{4}$European X-Ray Free-Electron Laser Facility, Holzkoppel 4, 22869 Schenefeld, Germany\\[0.5ex]
  $^{5}$Fakultät Physik/DELTA, TU Dortmund, 44221 Dortmund, Germany\\[0.5ex]
  $^{6}$The Hamburg Centre for Ultrafast Imaging, Luruper Chaussee 149, 22761 Hamburg, Germany\\[0.5ex]
  $^{7}$Deutsches Elektronen-Synchrotron DESY, Notkestr. 85, 22607 Hamburg, Germany\\[2ex]
  $^\ast$Corresponding author. Email: michelle.dargasz@uni-siegen.de\\
  $^\dagger$Corresponding author. Email: christian.gutt@uni-siegen.de\\[2ex]
}
\maketitle
\subsubsection*{This PDF file includes:}
Supplementary text\\
Figure S1, S2, S3, S4 and S5\\
Table S1, S2 and S3
\newpage
\renewcommand{\thefigure}{S\arabic{figure}}
\renewcommand{\thetable}{S\arabic{table}}
\setcounter{figure}{0}
\subsection*{Comparison of scattering strength of ferritin and crowders in the XPCS signal}
\begin{figure}[h]
\centering
\includegraphics[scale=.5]{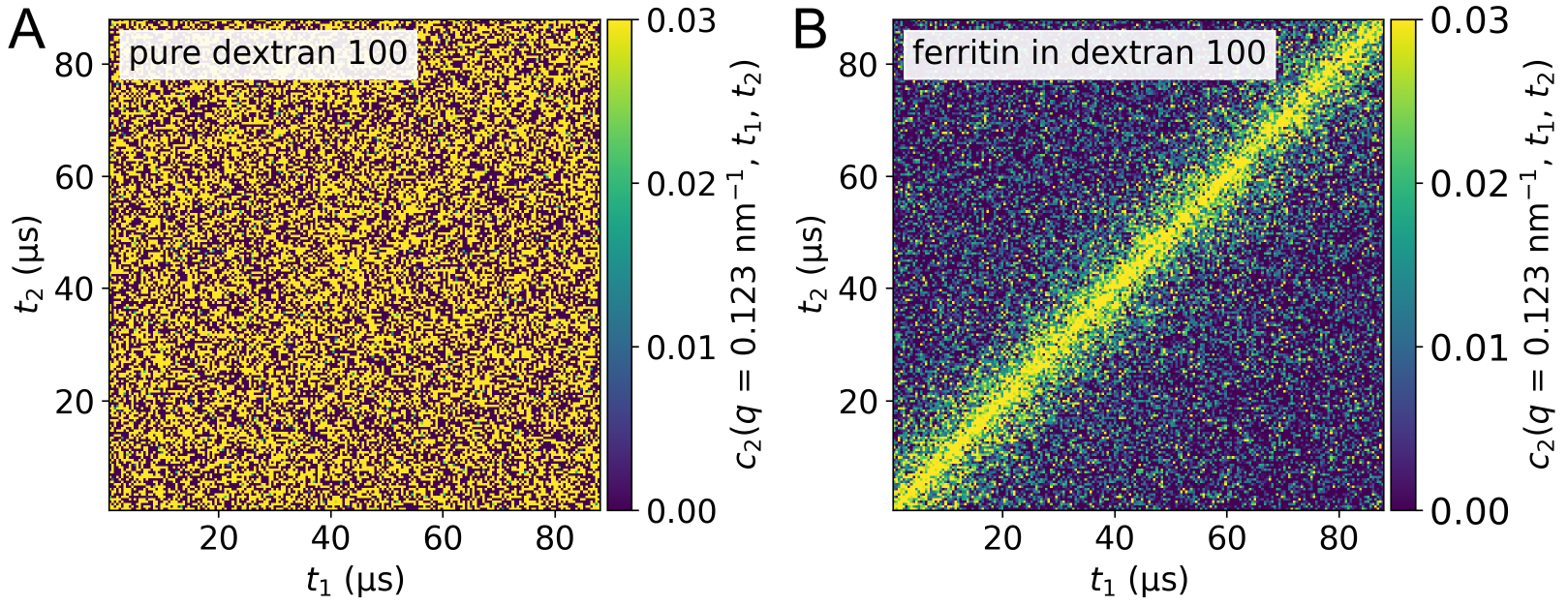}
\caption{Two-time correlation functions at $q = 0.123 \, \text{nm}^{-1}$ of (A) a pure 10 \%w/w dextran 100 solution and (B) ferritin in solution with 10 \%w/w dextran 100. }
\label{fig:ttc}
\end{figure}

In XPCS measurements, the contribution of each component to the measured correlation signal is weighted by its coherent scattering strength. Ferritin, due to its high electron density -- primarily from its iron-rich core -- and compact globular structure, exhibits a substantially stronger scattering signal per particle than polymeric crowders such as dextran 100. Although the crowder concentration is relatively high (10 - 35 \%w/w), these macromolecules possess low contrast relative to the aqueous solvent and are structurally diffuse, resulting in weak coherent scattering. As a result, the XPCS signal is dominated by ferritin, even at a moderate concentration of 55 mg/ml. This is consistent with the absence of a correlation signal in the pure crowder solution (Fig.~\ref{fig:ttc}A), reflecting the crowder’s low scattering power and confirming that the observed dynamics arise solely from ferritin, with negligible contribution from the crowders.

\subsection*{Crowder-specific parameters}
Table \ref{table:overlap} summarizes the relevant parameters for the polymeric crowders used in this study. Listed are the molecular weights ($M_{\mathrm{w}}$), the corresponding radii of gyration ($R_{\mathrm{g}}$) as reported in refs. \citeSI{Gra_JCollScie_1958,ranganathan2022ficoll_2}, and the overlap concentrations ($c^*$). For the dextran samples, $c^*$ was determined by identifying changes in the slope of the macroscopic viscosity, as obtained from measurements performed with a rotational viscometer. In the case of Ficoll, the overlap concentration was taken from the literature \citeSI{spencer2024concentration}.
\begin{table}[h!]
\caption{Molecular weights ($M_{\mathrm{w}}$), radii of gyration ($R_{\mathrm{g}}$) and overlap concentrations ($c^*$) of the polymeric crowders used in this study.}
\vspace{10pt}
\centering
\begin{tabular}{ |c|c|c|c| }
\hline
crowder & $M_{\mathrm{w}}$ (g/mol) & $R_{\mathrm{g}}$ (nm) & $c^*$ (\%w/w)\\
\hline 
Ficoll & 400 000 & 11.07 & 17 \\
dextran 40 & 40 000 & 6.2 & 12.75 \\
dextran 100 & 100 000 & 9.5 & 9.42 \\
dextran 500 & 500 000 & 20 & 4.97 \\
\hline
\end{tabular}
\label{table:overlap}
\end{table}
\subsection*{Limitations in $S(q)$ determination in case of binary mixtures}
To evaluate crowder-mediated protein-protein interactions, the structure factor of the pure ferritin-ferritin interactions in crowder solution would ideally be derived by dividing the back-\allowbreak ground-subtracted SAXS intensity of a concentrated protein solution ($I_{\text{hc}} - I_{\text{b}}$) by that of a dilute solution ($I_{\text{lc}} - I_{\text{b}}$), following:
\begin{equation*}
    S_{\text{ferr}-\text{ferr}}(q) \propto \frac{I_{\text{hc}}(q) - I_{\text{b}}(q)}{I_{\text{lc}}(q) - I_{\text{b}}(q)}
\end{equation*}
thereby isolating interparticle contributions from the single-particle form factor. However, in binary protein-crowder systems, where the two components exhibit distinct $q$-dependent scattering profiles, direct extraction of $S_{\text{ferr}-\text{ferr}}(q)$ becomes challenging. In such cases the total scattering intensity comprises the three contributions:
\begin{equation*}
I(q) \propto I_{\text{ferr}-\text{ferr}}(q) + I_{\text{ferr}-\text{cr}}(q) + I_{\text{cr}-\text{cr}}(q),
\end{equation*}
representing ferritin–ferritin correlations, ferritin–crowder cross-correlations, and crowder–\linebreak crowder interactions, respectively. Unlike pure systems, the cross-terms $I_{\text{ferr}-\text{cr}}$ cannot be removed through simple background subtraction, rendering it difficult to unambiguously isolate the ferritin-ferritin structure factor.  

It should be emphasized that the cross-correlations are particularly relevant in the analysis of the structure factor, as its interpretation relies on small deviations from unity. Even deviations of just a few percent can significantly affect the conclusions. In contrast, such minor deviations are not critical for the dynamic signal obtained from the XPCS measurements. Therefore, our analysis primarily focused on the more robust dynamic data, which are less susceptible to small uncertainties arising from cross-term contributions.

\begin{figure}[h]
\centering
\includegraphics[scale=.52]{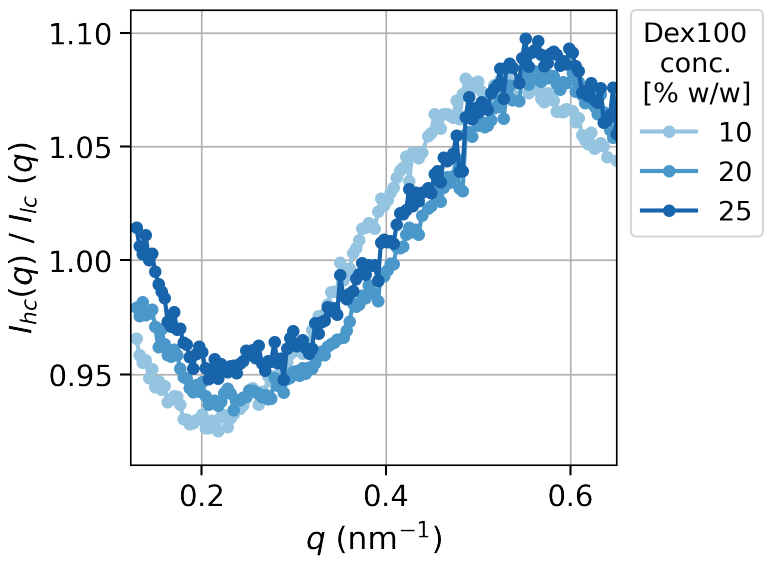}
\caption{Experimentally measured SAXS intensity of 55 mg/ml ferritin in solution with 10, 20 and 25 \%w/w dextran divided by SAXS intensity of 5 mg/ml ferritin in solution with the same dextran concentration.}
\label{fig:s(q)_diff}
\end{figure}

Fig.~\ref{fig:s(q)_diff} displays the ratio of the background-subtracted SAXS intensity of concentrated ferritin-crowder mixtures to that of a dilute ferritin solution at the same crowder  concentration. This quantity can be interpreted as an effective structure factor. The resulting profiles exhibit an increase at low $q$, consistent with the emergence of IRO. However, this effective structure factor inherently includes cross-correlations, which lead to deviations from the true ferritin-ferritin structure factor, $S_{\text{ferr}-\text{ferr}}(q)$, even when the crowder's intrinsic  scattering contribution is small compared to that of ferritin.

\subsection*{Modelling of $S(q)$ and $H(q)$}
The static structure factor $S$($q$) was calculated based on interparticle interactions by solving the Ornstein-Zernike equation with the mean spherical approximation closure (MSA). This is achieved with Python-based codes provided by the program \textit{jscatter} \citeSI{Bie_PLoS_2019}. 
Systems exhibiting  short-range attraction and long-range repulsion were modeled using a two-Yukawa potential of the form \citeSI{Liu_ChemPhys_2005}
\begin{equation}
\frac{V(r)}{k_{\mathrm{B}} \,T}=\begin{cases}
    -K_1 \frac{e^{-Z_1(r-1)}}{r} - K_2 \frac{e^{-Z_2(r-1)}}{r}     &\text{ for } r > 1 \\
    \;\;\infty   &\text{ for } 0 < r < 1 
    \end{cases}
    \label{eq:two-yukawa}
\end{equation}
where $Z_i$= 1/$\lambda_i$ denotes the inverse screening length, $K_i$ the potential strength in $k_{\mathrm{B}}\,T$ and $r$ is scaled to yield a hard-core diameter of 1.

For these calculations, the parameters of the repulsive term ($K_2$, $\lambda_2$) of the potential were held constant, while the attractive parameters $K_1$ and $scl_1$ were varied to fit the experimental data. In cases where the system exhibited primarily repulsive interactions, a single-Yukawa potential was used by setting $K_1=0$ and adjusting $K_2$ and $\lambda_2$. In both approaches, the ferritin radius and concentration were fixed, with a radius of 5.9 nm \citeSI{Wie_JApplCryst_2021_2} and ferritin concentration of 0.125 mmol/l.

\noindent The hydrodynamic function $H$($q$) was calculated using the $\delta\gamma$-expansion from Beenakker and Mazur \citeSI{Bee_PhysAStat_1983_2,Bee_PhysAStat_1984_2}. This formalism accounts for many-body hydrodynamic interactions in suspensions of spherical particles, using $S$($q$) as input. Following the implementation in \citeSI{Gen_PhysAStat_1991}, the $H$($q$) calculations were performed using \textit{jscatter}:
\begin{equation}
H(q) = H_{\text{d}}(q) + \frac{D_{\text{s}}(\Phi)}{D_0}
\end{equation}
where
\begin{equation}
H_{\text{d}}(q) = \frac{3}{2\pi}\int_0^{\infty} \text{d} R\,k \,\frac{\text{sin}^2(R\, k)}{(R\,k)^2(1+\Phi S_{\gamma}(R\,k))} \int_{-1}^1 \text{d}x (1-x^2) S(|q-k|-1)
\end{equation}
\begin{equation}
\frac{D_{\text{s}}(\Phi)}{D_0} = \frac{2}{\pi} \int_0^{\infty} \text{d}x \,\,\text{sinc}^2(x)(1+\Phi S_{\gamma}(x))^{-1}
\end{equation}
here $x = \text{cos}(q,k)$ represents the angle between the vectors $q$ and $k$, $\Phi$ is the protein volume fraction, and $R$ the protein radius. The function  $S_{\gamma}(x)$ is described in ref.  \citeSI{Gen_PhysAStat_1991}.

\subsection*{Interaction potentials derived from $D(q)$ analysis}
\begin{figure}[h]
\centering
\includegraphics[scale=.52]{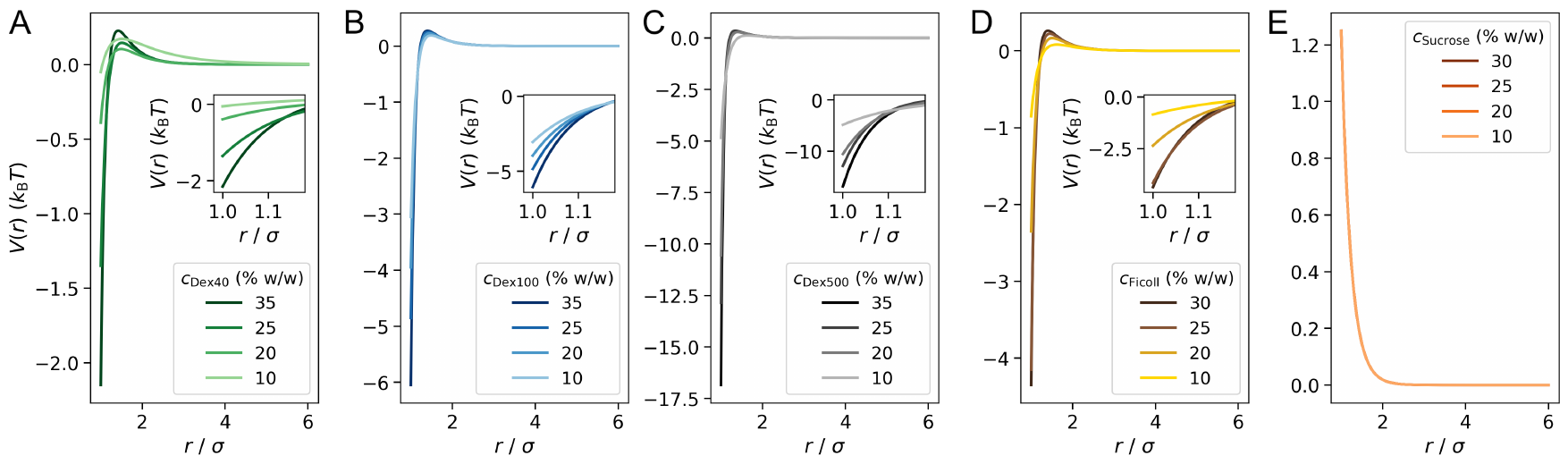}
\caption{Resulting two-Yukawa and single-Yukawa interaction potentials $V(r)$ of ferritin particles with diameter $\sigma$ = 11.8 nm in solution with (A) dextran 40, (B) dextran 100, (C) dextran 500, (D) Ficoll, and (E) sucrose. The insets highlight the varying potential minima.}
\label{fig:pot}
\end{figure}

The single-Yukawa and two-Yukawa interaction potentials $V(r)$, used to model the $D(q)$ of ferritin, are displayed in Fig.~\ref{fig:pot} for all crowder systems investigated. The corresponding parameters used to compute these potentials are summarized in  Table~\ref{table:params}. 

For sucrose solutions, we found that a single-Yukawa potential with fixed parameters across all crowder concentrations adequately describes the data, requiring only the diffusion coefficient $D_0(c)$ to be adjusted. As a result, the interaction potential shown in Fig.~\ref{fig:pot}(E) remains unchanged with varying sucrose concentration.
\begin{table}[h!]
\caption{Parameters used for modeling the $D(q)$ using a single-Yukawa and two-Yukawa interaction potential. $K_1$ accounts for the strength of the attractive part of the potential and $K_2$ for the repulsive part. The corresponding screening lengths are given by $\lambda_i$. $D_0(c_{\mathrm{cr}})$ refers to the diffusion constant of ferritin in crowder solutions at varying crowder concentrations in absence of all interactions. It should be noted that for the lowest dextran 40 concentration, a reduced repulsive interaction strength was required to achieve a satisfactory fit, as this concentration lies just below $c^*$ and thus close to the transition from a purely repulsive (single-Yukawa) to a weakly attractive (two-Yukawa) interaction potential. }
\vspace{10pt}
\centering
\begin{tabular}{ |c||c|c|c|c|c|c| }
\hline
crowder & conc. & $K_1$ & $\lambda_1$ & $K_2$  & $\lambda_2$ & $D_0(c_{\mathrm{cr}})$\\
 & (\%w/w) &$(k_{\mathrm{B}}T)$ &(nm) &$(k_{\mathrm{B}}T)$ &(nm) & (nm$^2$/$\upmu$m)\\
\hline 
sucrose & 10 &  &  & -1.25 & 3.43 & 19.7 $\pm$ 0.3\\
        & 20 &  &  & -1.25 & 3.43 & 12 $\pm$ 0.2\\
        & 25 &  &  & -1.25 & 3.43 & 8.9 $\pm$ 0.2\\
        & 30 &  &  & -1.25 & 3.43 & 6 $\pm$ 0.15\\
\hline 
Ficoll & 10 & 2 & 3.24 & -1.15 & 6 & 8.9 $\pm$ 0.3 \\ 
       & 20 & 3.5 & 2 & -1.15 & 6 & 1.9 $\pm$ 0.1 \\ 
       & 25 & 5.3 & 1.5 & -1.15 & 6 & 1.03 $\pm$ 0.03 \\ 
       & 30 & 5.5 & 1.33 & -1.15 & 6 & 0.37 $\pm$ 0.02 \\ 
\hline
dextran 40 & 10 & 0.7 & 4.8 & -0.65 & 17.14 & 5.9 $\pm$ 0.2 \\
          & 20 & 1.54 & 3.43 & -1.15 & 6 & 1.58 $\pm$ 0.03 \\
          & 25 & 2.5 & 2.4 & -1.15 & 6 & 0.48 $\pm$ 0.02 \\
          & 35 & 3.3 & 1.71 & -1.15 & 6 & 0.245 $\pm$ 0.015 \\
\hline
dextran 100 & 10 & 4.2 & 1.76 & -1.15 & 6 & 4.8 $\pm$ 0.3 \\
           & 20 & 5.1 & 1.5 & -1.15 & 6 & 1.22 $\pm$ 0.05 \\
           & 25 & 6 & 1.33 & -1.15 & 6 & 0.74 $\pm$ 0.025 \\
           & 35 & 7.2 & 1.2 & -1.15 & 6 & 0.3 $\pm$ 0.035 \\
\hline
dextran 500 & 10 & 6 & 2 & -1.15 & 6 & 5.3 $\pm$ 0.15 \\
           & 20 & 11.7 & 1.04 & -1.15 & 6 & 1.23 $\pm$ 0.045 \\
           & 25 & 14 & 0.86 & -1.15 & 6 & 0.8 $\pm$ 0.05 \\
           & 35 & 18 & 0.8 & -1.15 & 6 & 0.155 $\pm$ 0.025 \\
\hline
\end{tabular}
\label{table:params}
\end{table}

Upon examining the collective diffusion coefficient  of ferritin in dextran 500 solutions, it becomes evident that, particularly at 10 \%w/w and 35 \%w/w, the characteristic change in slope of $D(q)$ at low $q$ -- as observed for  Ficoll and other dextran molecular weights -- is absent. This raises the question of whether, in these cases, $D(q)$ could be described using a purely attractive particle interaction potential. To test this, the same fitting procedure applied to the single- and two-Yukawa models was employed using a purely attractive sticky hard sphere potential, given by \citeSI{menon1991new,sztucki2006kinetic}
\begin{equation}
V(r)=\begin{cases}
    \;\;\infty   & 0< r < \sigma \\
    - U_0      & \sigma \leq r \leq \sigma+\Delta \\
    \;\;\;0   &\sigma +\Delta < r  
    \end{cases}
\end{equation}
where $U_0$ is the potential well depth, $\Delta$ the width of the square well and $\sigma$ the ferritin diameter. 

\begin{figure}[h]
\centering
\includegraphics[scale=.52]{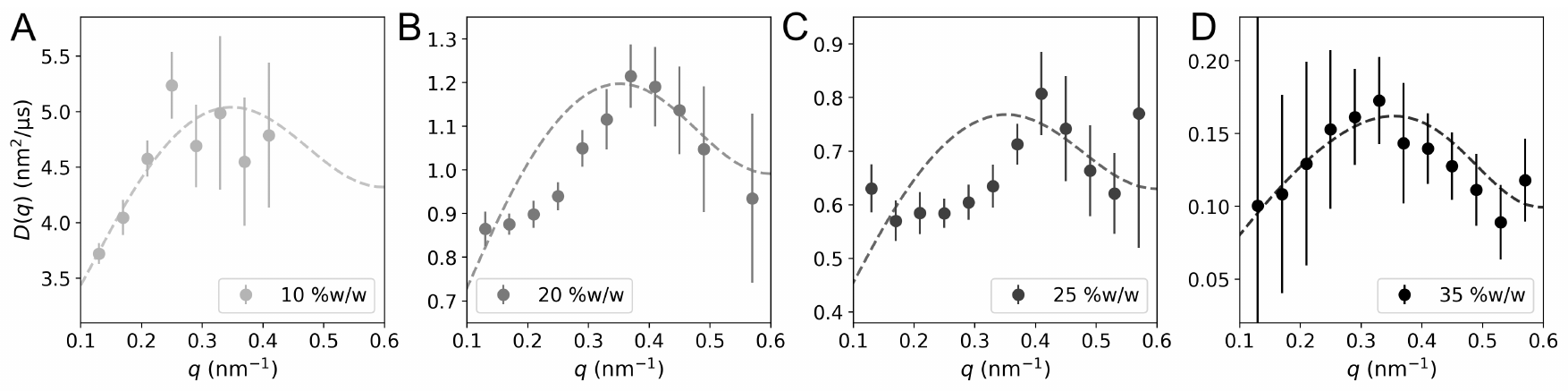}
\caption{Sticky hard sphere fits to the collective diffusion coefficient $D(q)$ of ferritin in solutions with dextran 500 at concentrations of (A) 10 \%w/w, (B) 20 \%w/w, (C) 25 \%w/w and (D) 35 \%w/w. Parameters in Table ~\ref{table:params_sticky}. }
\label{fig:pot_sticky}
\end{figure}

The results for all four concentrations (10 – 35 \%w/w) are shown in Fig.~\ref{fig:pot_sticky} with the corresponding fitting parameters listed in Table~\ref{table:params_sticky}.
At concentrations where no change in slope at low $q$ is observed (10 \%w/w and 35 \%w/w), the sticky hard sphere potential provides a reasonable description of the $D(q)$ behavior. However, it becomes clear that as soon as a low-$q$ slope change emerges -- which is also observed for Ficoll and the other dextran molecular weights -- the purely attractive sticky hard sphere potential is no longer sufficient. In these cases, an additional long-range repulsive component must be included alongside the short-range attraction to accurately describe the observed diffusion profiles.

The behavior of ferritin in dextran 500 may therefore be attributed either to a non-linear concentration dependence of the interaction potential - becoming more repulsive at intermediate concentrations and more attractive at higher ones, or simply to the limited $q$-range, which may have prevented the observation of an upturn at lower $q$.

\begin{table}[h!]
\caption{Parameters used for modeling the $D(q)$ using a sticky hard sphere interaction potential. $U_0$ represents the potential well depth, $\Delta$ indicates the width of the square well and $D_0(c_{\mathrm{cr}})$ refers to the protein diffusion constant at varying dextran 500 concentration in absence of all interactions. }
\vspace{10pt}
\centering
\begin{tabular}{ |c|c|c|c| }
\hline
conc. (\%w/w) & $\Delta$ (nm) & $U_0$ $(k_{\mathrm{B}}T)$ & $D_0(c)$ (nm$^2$/$\upmu$m)\\
\hline 
 10 & 0.8 & 2.4 & 5.4 $\pm$ 0.12\\
 20 & 0.8 & 2.6 & 1.255 $\pm$ 0.05\\
 25 & 0.8 & 2.65 & 0.8 $\pm$ 0.06\\
 35 & 0.8 & 3 & 0.14 $\pm$ 0.02\\
\hline 
\end{tabular}
\label{table:params_sticky}
\end{table}

\subsection*{Slow dynamics of ferritin in highly concentrated dextran 500 solutions}
\begin{figure}[h]
\centering
\includegraphics[scale=.4]{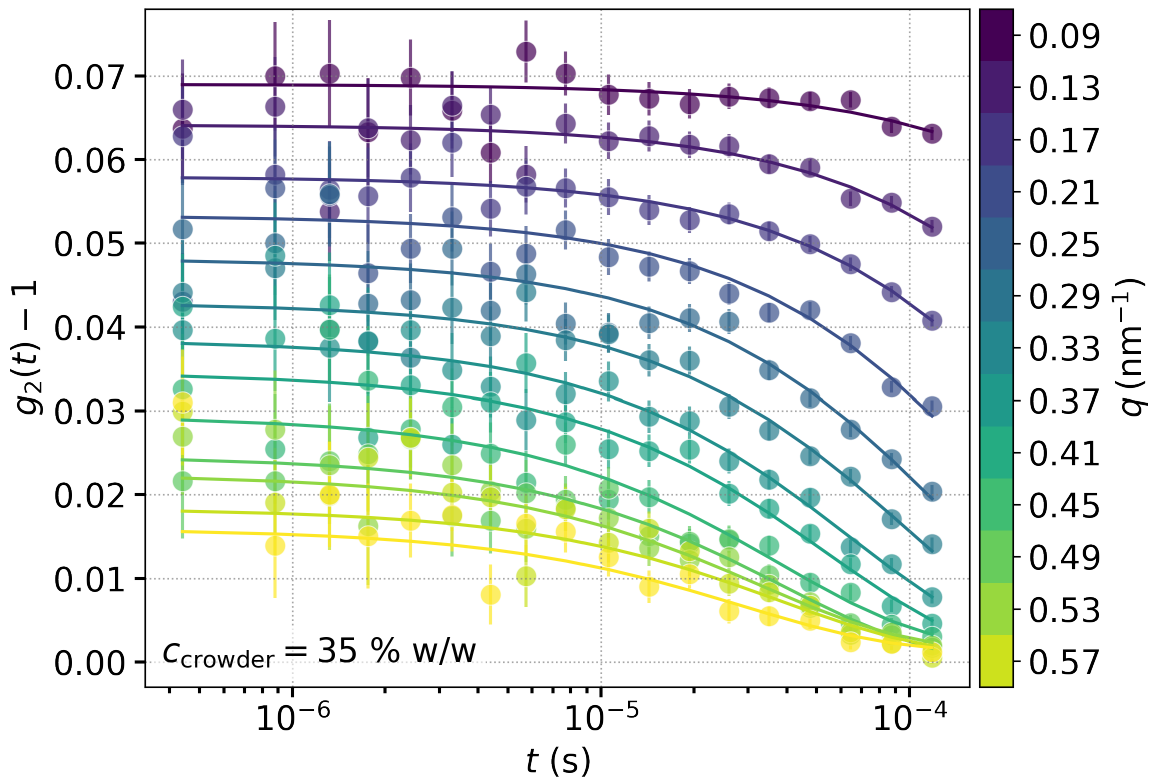}
\caption{$g_2(q,t)$ functions for ferritin in solution with 35 \%w/w dextran 500, showing markedly slow dynamics at low $q$-values consistent with the calculated slow fluctuations of IRO. }
\label{fig:g2_dex500}
\end{figure}

The $q$-dependent $g_2(t)$ functions for ferritin in 35 \%w/w dextran 500 solution, shown in Fig.~\ref{fig:g2_dex500}, reveal markedly slow protein dynamics. In particular the $g_2(t)$ functions at low $q$-values show no visible decay within the experimentally accessible time window. This behavior is consistent with the estimated long lifetimes of protein complexes associated with the IRO in dextran (see Fig.~\ref{fig:s(q)}).
\subsection*{Estimation of depletion layer thickness and reduced viscosity within the depletion layer}
The depletion layer thickness $\delta_{\text{s}}$ surrounding ferritin was estimated using the experimentally determined correlation length $\xi$, following the approach described in refs.   \citeSI{Zos_pnas_2020_2,fleer2003mean_2,tuinier2002interaction_2,hanke1999polymer_2}
\begin{align}
    \delta_{\text{s}} &= R \left(1 + 3\frac{\delta}{R}+2.273\left(\frac{\delta}{R}\right)^2-0.0975 \left(\frac{\delta}{R}\right)^3\right)^{1/3} - R,
    \label{eq:depl_thick}
\end{align}
where $\delta^{-2} = \delta_0^{-2} + \xi^{-2}$. $\delta_0$ is the depletion layer thickness in the dilute regime, approximated by the polymer's radius of gyration $R_{\mathrm{g}}$ (see Table~\ref{table:overlap}) and $R$ corresponds to the radius of ferritin. 

To estimate the microscopic viscosity $\eta_{\text{micro}}$ within the depletion layer, the model of Tuinier et al.  \citeSI{tuinier2006depletion_2} was employed with
\begin{align}
    \label{eq:micro_visc}
    \eta_{\text{micro}}&= \eta_{\text{s}} \frac{Q(\lambda,\epsilon)}{Z(\lambda,\epsilon)},
\end{align}
where $\eta_s$ is the solvent viscosity, and
$\lambda = \frac{\eta_{\text{s}}}{\eta_{\text{macro}}}$ is the ratio of solvent to macroscopic viscosity $\eta_{\text{macro}}$ of the crowder solution.  The dimensionless depletion layer thickness is given by $\epsilon = \frac{\delta_{\text{s}}}{R}$. The functions  $Q(\lambda,\epsilon)$ and $Z(\lambda,\epsilon)$ are defined as:
\begin{align}
    Q(\lambda,\epsilon) &= 2(2+3\lambda)(1+\epsilon)^6-4(1-\lambda)(1+\epsilon) \notag \\
    Z(\lambda,\epsilon) &= 2(2+3\lambda)(1+\epsilon)^6-9\left(1-\frac{1}{3}\lambda -\frac{2}{3} \lambda^2\right)(1+\epsilon)^5 \notag\\&+10(1-\lambda)(1+\epsilon)^3-9(1-\lambda)(1+\epsilon)+4(1-\lambda)^2.
\end{align}

\bibliographystyleSI{pnas-new}
\bibliographySI{main}
\end{document}